\documentclass[epj]{svjour}
\usepackage{graphics}
\usepackage{appendix}
\usepackage[dvipsnames]{xcolor}
\usepackage[T1]{fontenc}
\usepackage{graphicx}
\usepackage{amsmath}
\usepackage{amssymb}
\usepackage{xspace}
\usepackage[numbers,sort&compress]{natbib}
\usepackage{subfig}
\usepackage[shortlabels]{enumitem}
\usepackage{float}
\usepackage{pstricks}
\usepackage{graphicx}
\usepackage{xspace}
\usepackage{comment}
\usepackage{lipsum}
\usepackage{multirow}
\usepackage{placeins}
\usepackage{braket}
\usepackage[compat=1.1.0]{tikz-feynman}
\usepackage{contour}
\usepackage{tikz}
\usepackage{placeins}
\usepackage{ulem}
\usepackage[colorlinks=true, linkcolor=blue, citecolor=blue, urlcolor=blue]{hyperref}

\newcommand{\calM}{\mathcal{M}}
\newcommand{\phiphantK}{ \varphi_{{k}^{\phantom{\prime}}}}

\newcommand{\phiphantJ}{ \varphi_{{j}^{\phantom{\prime}}}}

\newcommand{\as}{\alpha_\text{s}}

\newcommand{\LO}{\text{LO}\xspace}

\newcommand{\EW}{\ensuremath{\text{EW}}\xspace}

\newcommand{\NVI}{\ensuremath{\text{NLO}_{\text{VI}}~\EW}\xspace}
\newcommand{\NLL}{\ensuremath{\text{NLL}_{\text{VI}}~\EW}\xspace}
\newcommand{\NLLp}{\ensuremath{\text{NLL'}_{\text{VI}}~\EW}\xspace}

\newcommand{\SSSC}{\ensuremath{\text{S-SSC}}\xspace}
\newcommand{\COLL}{\ensuremath{\text{C}}\xspace}

\newcommand{\LL}{\ensuremath{\text{LL}}\xspace}
\newcommand{\NLLai}{\ensuremath{\text{NLL a.i.}}\xspace}
\newcommand{\NLLad}{\ensuremath{\text{NLL a.d.}}\xspace}
\newcommand{\ai}{\ensuremath{\text{a.i.}}\xspace}
\newcommand{\ad}{\ensuremath{\text{a.d.}}\xspace}

\newcommand{\nll}{\text{NLL}\xspace}

\newcommand{\NLLpone}{\ensuremath{\text{NLL'}_{\text{VI}}^{\,\,(1)}~\EW}\xspace}
\newcommand{\NLLtwo}{\ensuremath{\text{NLL}_{\text{MR}}^{\,\,(2)}~\EW}\xspace}
\newcommand{\nNLO}{\ensuremath{\text{nNLO}_{\text{MR}}^{\,\,(2)}~\EW}\xspace}

\newcommand{\Sherpa}{{\rmfamily\tt Sherpa}\xspace}

\newcommand{\OpenLoops}{{\rmfamily\tt OpenLoops}\xspace}

\newcommand{\Rivet}{{\rmfamily\tt Rivet}\xspace}

\newcommand{\rR}{\mathrm{R}}

\newcommand{\mur}{\mu_{\rR}}

\newcommand{\muew}{\ensuremath{\mu_{\rm PR}}\xspace}
\newcommand{\muren}{\ensuremath{\mu_{\rm R}}\xspace}

\newcommand{\ord}{\mathcal{O}}

\newcommand{\beqar}{\begin{eqnarray}}
\newcommand{\eeqar}{\end{eqnarray}}
\newcommand{\beq}{\begin{equation}}
\newcommand{\eeq}{\end{equation}}
\newcommand{\bit}{\begin{itemize}}
\newcommand{\eit}{\end{itemize}}

\def\setrelwidth{0.40}

\usepackage[left,mathlines]{lineno} % line numbering
\usepackage{soul} %text highlighting
\begin{document}
\title{Logarithmic EW corrections at two-loop}
\author{J. M. Lindert\inst{1}, L. Mai\inst{2} % etc
}                     % Do not remove
\institute{
\inst{1} Department of Physics and Astronomy, University of Sussex, Brighton BN1 9QH, UK\\
\inst{2} Dipartimento di Fisica, Universit\`a di Genova and INFN, Sezione di Genova, Via Dodecaneso 33, 16146, Italy 
} 

\date{Received: date / Revised version: date}
\abstract{
We present the implementation of next-to-next-to-leading order (NNLO) electroweak
(EW) virtual corrections at next-to-leading logarithmic (NLL) accuracy in the
amplitude generator \OpenLoops. The implementation covers the automated computation of processes
involving massless fermions and transversely polarised vector bosons.
For energies above the EW scale, logarithmic EW corrections are strongly enhanced in
the tails of kinematic distributions of key LHC processes, reaching several tens of
percent at NLO and several percent at NNLO. The two-loop implementation is validated
against analytical results from the literature. We present phenomenological results for
representative LHC processes and discuss the role of two-loop EW corrections in
reducing theoretical uncertainties from missing higher-order contributions. 
%
%\PACS{
%      {PACS-key}{discribing text of that key}   \and
%      {PACS-key}{discribing text of that key}
%     } % end of PACS codes
} %end of abstract
\maketitle
%

%%%%%%%%%%%%%%%%%%%%%%%%%%%%%%%%%%%%%%%%%%%%%%%%%%%%%%%%
\section{Introduction} \label{sec:intro}

Electroweak (EW) radiative corrections to high-energy processes are characterised by the presence of logarithmic terms of the ratio $Q^2/M^2$, where $Q$ denotes the typical scattering energy and $M$ is the electroweak gauge-boson mass scale, with $M = M_W \sim M_Z$~\cite{Fadin:1999bq,Kuhn:1999nn,Denner:2000jv,Denner:2001gw,Ciafaloni:2000df,Baur:2006sn,Mishra:2013una}. These logarithmic corrections affect all reactions involving electroweakly interacting particles at energy scales $Q \gg M$. At the LHC and future lepton colliders, they can amount to several tens of percent at next-to-leading order (NLO) and still several percent at next-to-next-to-leading order (NNLO). This class of corrections therefore plays a crucial role in the interpretation of precision measurements and in backgrounds for searches at the LHC and future high-energy colliders.

At sufficiently high $Q$, N$^k$LO EW corrections generate a logarithmic structure
\begin{equation}\label{eq:log_tower}
\left(\frac{\alpha}{4\pi}\right)^k \log^n\left(\frac{Q^2}{M^2}\right), \qquad 0\leq n\leq 2k\,,
\end{equation}
where the power $n=2k$ corresponds to leading logarithms (LL) and $n=2k-1$ to next-to-leading logarithms (NLL). These logarithmic corrections have a twofold origin: they arise as $\log(Q^2/\mur^2)$ from the renormalisation of ultraviolet singularities at the scale $\mur \sim M$, and as mass singularities $\log(Q^2/M^2)$ from the interaction of external on-shell particles with soft and/or collinear gauge bosons.

The UV and IR origin of EW logarithmic corrections enables a process-independent treatment and the derivation of universal correction factors in terms of kinematics and electroweak quantum numbers of the external particles. At one loop, a general expression capturing the LL and NLL contributions to arbitrary processes was derived in Refs.~\cite{Denner:2000jv, Denner:2001gw, Pozzorini:2001rs}. The algorithmic construction of these corrections has since been implemented in several automated tools~\cite{Chiesa:2013yma,Bothmann:2020sxm,Pagani:2021vyk, Lindert:2023fcu, Mai:2024fsk}. In particular, the implementations in {\tt Sherpa}~\cite{Bothmann:2020sxm}, {\tt MadGraph5\_aMC@NLO}~\cite{Pagani:2021vyk,Pagani:2023wgc,ElFaham:2024egs} and \OpenLoops~\cite{Lindert:2023fcu} are fully general and applicable to any hard SM process in the regime where all kinematic invariants are of the same order and much larger than the EW scale. The \OpenLoops implementation \cite{Lindert:2023fcu} additionally supports a generalisation to resonant processes involving the decay of heavy particles.

Beyond one loop, the properties of electroweak logarithmic corrections have been studied via two complementary approaches. On one hand, the infrared evolution equation (IREE) method~\cite{Fadin:1999bq,Melles:2000gw,Melles:2000ia,Melles:2001mr,Melles:2001dh,Kuhn:1999nn,Kuhn:2001hz} resums higher-order terms under the assumption that the theory can be split into two regimes with exact gauge symmetry: a symmetric $SU(2) \otimes U(1)$ phase for scales $Q \geq \mu_{\text{IREE}} \geq M$, and an unbroken $U(1)_{\rm em}$ phase for $\mu_{\text{IREE}} \leq M$. Alternatively, the resummation of EW Sudakov logarithms has been formulated within soft-collinear effective theory (SCET)~\cite{Chiu:2007yn,Chiu:2007dg,Chiu:2009yz,Fuhrer:2010eu}, which allows for the systematic inclusion of additional power corrections, and phenomenological results have been presented for several LHC processes~\cite{Chiu:2008vv,Fuhrer:2010vi,Assi:2020lbs,Denner:2024yut}. On the other hand, explicit diagrammatic computations in the broken EW Standard Model, where effects of symmetry breaking are automatically included, have been performed for specific processes in Refs.~\cite{Melles:2000ed,Hori:2000tm} and in a process-independent manner in Refs.~\cite{Beenakker:2001kf,Denner:2003wi,Denner:2006jr,Denner:2008yn}. These explicit results confirmed the exponentiation of both LL and the full set of NLL terms as predicted by the IREE and SCET approaches.

In this paper we present an implementation for the automated computation of two-loop EW corrections in logarithmic approximation within the amplitude generator \OpenLoops~\cite{Buccioni:2019sur}, applicable to processes involving an arbitrary number of massless fermions and transversely polarised gauge bosons. The generalisation to longitudinal gauge bosons, massive fermions and scalars will be presented in a future publication. 
Our implementation combines two complementary strategies: the angular-dependent two-loop logarithms are computed following the diagrammatic approach in the broken EW Standard Model of Refs.~\cite{Denner:2003wi,Denner:2006jr}, while the angular-independent logarithms are constructed based on standard counterterms and wavefunction corrections ensuring consistency with the all-order resummation results of Refs.~\cite{Melles:2000gw,Melles:2001mr}. In both methods, infrared singular photon/QED contributions are regularised in mass regularisation (MR). The implementation builds upon and extends the automated one-loop EW Sudakov corrections of Ref.~\cite{Lindert:2023fcu}, exploiting the same pseudo-\linebreak counterterm approach for the efficient generation of the required $SU(2)$-correlated Born amplitudes. We investigate the scope of our implementation for a set of representative LHC processes including $V$+jet, $V$+2\,jets, $VV$+jet and $VVV$ production.

The structure of this article is as follows. In Section~\ref{sec:theory} we revisit the structure of one- and two-loop EW Sudakov logarithms and their all-order resummation. In Section~\ref{sec:implementation} we describe the implementation in \OpenLoops, focussing on the pseudo-counterterm approach and its extension to two loops. Numerical results are presented in Section~\ref{sec:results}, and we conclude in Section~\ref{sec:conclusions}. Further details and validation material are collected in the Appendices.

%%%%%%%%%%%%%%%%%%%%%%%%%%%%%%%%%%%%%%%%%%%%%%%%%%%%%%%%

%%%%%%%%%%%%%%%%%%%%%%%%%%%%%%%%%%%%%%%%%%%%%%%%%%%%%%%%
\section{Logarithmic EW corrections} \label{sec:theory}
%%%%%%%%%%%%%%%%%%%%%%%%%%%%%%%%%%%%%%%%%%%%%%%%%%%%%%%%

%%%%%%%%%%%%%%%%%%%%%%%%%%%%%%%%%%%%%%%%%%%%%%%%%%%%%%%%
\subsection{Notation and Logarithmic Approximation} 
%%%%%%%%%%%%%%%%%%%%%%%%%%%%%%%%%%%%%%%%%%%%%%%%%%%%%%%%

For the perturbative expansion of matrix elements in $\alpha =e^2/(4\pi)$, where $e$ is the electromagnetic coupling constant, we write
\begin{equation}\label{eq:pert_exp}
\mathcal{M} = \sum_{k=0}^{\infty} \left(\frac{\alpha}{4\pi}\right)^k \mathcal{M}_k.
\end{equation}
We use a convention similar to Refs.~\cite{Denner:2000jv,Denner:2006jr}, and in the following consider a $n \to 0$  process
\begin{equation} \label{process_convention}
            \varphi_1(p_1) \ldots \varphi_n(p_n) \to 0\,,
        \end{equation}
where predictions for the corresponding physical process are obtained via crossing symmetries. 
Here $\varphi_i$ 
can be any chiral massless fermion $f_{j,\sigma}^{\kappa}$, $\bar{f}_{j,\sigma}^{\kappa}$ (with $f = Q,L$, chirality $\kappa=L,R$, weak isospin $\sigma=\pm$, generation $j=1,2,3$) or transverse gauge boson $V_{\rm T}=A_{\rm T}, Z_{\rm T}, W^{\pm}_{\rm T}$.

We consider processes whose Born-level amplitude \linebreak $\mathcal{M}_0$ 
is not mass-suppressed, i.e.\ $\mathcal{M}_0 \sim E^d$  
for $E=\sqrt{s} \gg m_{W}$, with $d$ the mass dimension of the matrix element, and where all kinematic invariants are much larger than the gauge-boson masses,
\begin{align}\label{la}
|r_{jk}|&=|(p_j+p_k)^2| \nonumber \\
&\approx 2|p_jp_k| \sim Q^2\gg m_{W}^2,    \hspace{0.5 cm} j \neq k.
\end{align}
%Here, $m_j$ and $\lambda$ are respectively the fictitious fermion and photon masses employed for the regularisation of infrared singular photon/QED contributions within MR.

In the regime of Eq.~\eqref{la}, the one- and two-loop EW corrections are dominated by logarithmic terms of the form~\cite{Kuhn:1999de,Fadin:1999bq,Denner:2000jv,Pozzorini:2001rs,Denner:2003wi,Denner:2006jr,Denner:2008yn}
\begin{align}\label{eq:generic_oneloop}
\mathcal{M}_1^{
} &= (\hat\delta_1^{\rm{LL}} + \hat\delta_1^{\rm{NLL}}) \mathcal{M}_0^{
}   
\sim \left(\frac{{\alpha}}{4\pi}\right)  \left[L^2 + (a+l_{jk})L  \right] E^d\,,
\\\label{eq:generic_twoloop}
\mathcal{M}_2^{
} &= (\hat\delta_2^{\rm{LL}} + \hat\delta_2^{\rm{NLL}}) \mathcal{M}_0^{
}   
\sim \left(\frac{{\alpha}}{4\pi}\right)^2  \left[L^4 + (b+l_{jk})L^3 \right] E^d\,, 
\end{align}
where $\hat\delta_k^{\rm{LL}}$ and $\hat\delta_k^{\rm{NLL}}$ are the $\mathcal{O}(\alpha^k)$ leading- and next-to-leading logarithmic correction operators, $a$ and $b$ are constants\footnote{These constants are in general complex numbers due to the presence of the additional $i\pi$ factors coming from analytical continuation of $\log(-r_{jk})$. See Ref. \cite{Pagani:2021vyk,Lindert:2023fcu} for more details for relevance within EW Sudakov corrections.}, and the angular-independent and angular-dependent logarithms are defined as
\begin{align}\label{eq:logs_def_generic}
L &\equiv L(Q,m_i^2):= \log{\frac{Q^2}{m_i^2}}, \\
 l_{jk}&\equiv l(r_{jk},Q)= l_{jk}(Q):= \log{\frac{|r_{jk}|}{Q^2}}.
\end{align}
These logarithmic corrections are universal and factorise with respect to the corresponding Born or $SU(2)$-correlated Born amplitudes, enabling a process-independent implementation. The two-loop structure in Eq.~\eqref{eq:generic_twoloop} is consistent with the all-order exponentiation of EW Sudakov logarithms established via  IREE~\cite{Fadin:1999bq,Melles:2000gw,Melles:2001mr} and SCET~\cite{Chiu:2007yn,Chiu:2007dg,Chiu:2009yz}: at NLL accuracy, the angular-independent corrections exponentiate into per-leg Sudakov form factors, while the angular-dependent soft corrections exponentiate in the space of $SU(2)$-correlated amplitudes. The explicit two-loop coefficients in Eq.~\eqref{eq:generic_twoloop} correspond to the $\mathcal{O}(\alpha^2)$ expansion of these resummed expressions, as we discuss in detail in Section~\ref{sec:allorders}.

Not covered by the logarithmic approximation (LA) defined by Eqs.~\eqref{eq:generic_oneloop}--\eqref{eq:generic_twoloop} are mass-suppressed logarithmic contributions of the form $\alpha^k m^n_{W} E^{d-n}L^p$ with $n > 0$, as well as NNLL terms with $n=0$ and $p \le 2k-2$ .

%%%%%%%%%%%%%%%%%%%%%%%%%%%%%%%%%%%%%%%%%%%%%%%%%%%%%%%%
\subsection{EW Sudakov corrections at all orders} \label{sec:allorders}
%%%%%%%%%%%%%%%%%%%%%%%%%%%%%%%%%%%%%%%%%%%%%%%%%%%%%%%%

The properties of electroweak logarithmic corrections at all orders in $\alpha$ have been investigated by means of the IREE method~\cite{Fadin:1999bq,Melles:2000gw,Melles:2000ia,Melles:2001mr,Melles:2001dh}. 
The IREE approach exploits the fact that in the high-energy regime, where all invariants are much larger than the electroweak scale, the virtual corrections can be described by a generalised renormalisation group equation with infrared-singular anomalous dimensions.
The key assumption underlying the IREE treatment of the electroweak theory is
that the regime above the electroweak scale $M$ can be described by a symmetric $SU(2) \otimes U(1)$ theory, while the regime below $M$ is governed by an unbroken $U(1)_{\rm em}$, with appropriate matching conditions at the scale $\mu = M$.

\subsubsection*{Angular-independent NLL resummation}

In this framework, the all-order angular-independent (\ai) resummed amplitude for the process~\eqref{process_convention} involving $n$ external massless fermions or transversely polarised gauge bosons can be written in the factorised form\footnote{ In the case of external neutral vector bosons $V_N=A,Z$, one has to appropriately include mixing effects in Eq. \eqref{eq:resum_master} as discussed in Ref.~\cite{Melles:2000gw}. Indeed, for transverse degrees of freedom the corrections do not factorise with respect to the physical Born amplitude but rather with respect to the amplitudes containing the fields in the unbroken phase.}~\cite{Fadin:1999bq,Melles:2000gw,Melles:2001mr}
\begin{equation}\label{eq:resum_master}
\mathcal{M}^{\rm RG\, \ai} = \prod_{j=1}^{n} \Delta_j(s,M^2) \cdot \mathcal{M}_0 \,,
\end{equation}
where the Sudakov form factor $\Delta_j$ associated with each external leg $j$ exponentiates the angular-independent logarithmic corrections, while the angular-dependent soft-collinear contributions exponentiate via the generalisation of Eq.~\eqref{eq:resum_master} to a matrix structure in the space of $SU(2)$-correlated amplitudes, as discussed below.

More explicitly, the all-order angular-independent correction to the amplitude can be cast in the form
\begin{equation}\label{eq:resum_allorder}
\mathcal{M}^{\rm RG\, \ai} = \exp\left(-\sum_{j=1}^n  \delta^{\rm a.i.}_j\right)  \cdot \mathcal{M}_0 \,,
\end{equation}
where the exponent receives contributions from both the LL (soft-collinear) and angular-independent NLL (collinear and running of the gauge couplings) sectors.

At LL the IREE yields an exponential Sudakov form factor for each external particle $\varphi_j$ of the form
\begin{equation}\label{eq:LL_resum}
\delta^{\rm LL}_{\varphi_j}(s,M^2) = \frac{1}{2}\sum_a  C^{a}_{\varphi_j} \frac{\alpha_a}{4\pi} L^2(Q,M) ,
\end{equation}
where the sum runs over the gauge-group factors $a = \{SU(2)_L, U(1)_Y\}$, and
\begin{equation}\label{eq:casimirs}
C^{SU(2)}_{\varphi_j} = C_{\varphi_j}^{(2)} = T_{\varphi_j}(T_{\varphi_j}+1), \qquad C^{U(1)}_{\varphi_j} = \frac{Y_{\varphi_j}^2}{4} \,,
\end{equation}
are the $SU(2)$ Casimir eigenvalue and the squared hypercharge of the particle $\varphi_j$, respectively. Here $\alpha_a$ denotes the coupling constant of the gauge group $a$, with $\alpha_{SU(2)} = \alpha/s^2_{\rm w}$ and $\alpha_{U(1)} = \alpha/c^2_{\rm w}$, where $s_{\rm w}$ and $c_{\rm w}$ are the sine and cosine of the weak mixing angle. Note that the argument of the logarithm $L$ in~\eqref{eq:LL_resum} involves the common mass scale $M$ of the EW gauge bosons, reflecting the assumption of an approximately degenerate $SU(2) \otimes U(1)$ symmetric theory above $M$. In the following we suppress this explicit dependence of the logarithms. 

At NLL accuracy, the IREE approach of Refs.~\cite{Melles:2000gw,Melles:2001mr} yields additional single-logarithmic corrections that also exponentiate. These subleading contributions originate from: (i) collinear emission from external lines, described by the virtual parts of the Altarelli-Parisi splitting functions, and (ii) the running of the gauge couplings, which at higher orders gets folded with the LL Sudakov corrections.

The NLL contribution to the Sudakov form factor for each external particle $\varphi_j$ takes the form
\begin{equation}\label{eq:NLL_resum}
\Delta^{\NLLai}_{\varphi_j}(s,M^2) = \exp\left[-\sum_a \gamma^{a}_{\varphi_j} \frac{\alpha_a}{4\pi} L \right] ,
\end{equation}
where $\gamma^{a}_{\varphi_j}$ are the collinear anomalous dimensions associated with the external particle $\varphi_j$ for each gauge group factor $a$. These anomalous dimensions are obtained from the virtual parts of the regularised Altarelli-Parisi splitting functions~\cite{Melles:2000gw} (in the symmetric phase) as
\begin{align}\label{eq:anom_dim_fermion}
\gamma^{a}_{f} &= - \frac{3}{2} C^{a}_{f} \,,&\\[4pt]
\label{eq:anom_dim_gauge}
\gamma^{a}_{V_{\rm T}} &= 
\begin{cases}
-\dfrac{b^{(1)}_{SU(2)}\, s^2_{\mathrm{w}}}{2} & a = SU(2) \\[10pt]
-\dfrac{b^{(1)}_{U(1)}\, c^2_{\mathrm{w}}}{2}  & a = U(1)
\end{cases}
\end{align}
for massless fermions and transverse gauge bosons respectively. Here $C^{a}_f$ is the fermionic Casimir eigenvalue and $b^{(1)}_a$ is the one-loop $\beta$-function coefficient of the gauge group $a$, related to the standard one-loop $\beta$-function via $\mu \frac{d\alpha_a}{d\mu} = -\frac{b^{(1)}_a \alpha_a^2}{2\pi} + \mathcal{O}(\alpha_a^3)$.

If the gauge couplings $\alpha_a$ entering the Born subamplitude $\mathcal{M}_0$  are renormalised at the hard scale $\mur=Q$, their running doesn't yield any additional contribution and Eq.~\eqref{eq:NLL_resum} is the total NLL correction. On the contrary, on-shell renormalisation at $\mur=m_W$ introduces additional NLL logarithmic contributions stemming from the running of the gauge couplings from $\mur$ itself to $Q$; they can be included by performing the following replacements
\begin{equation}\label{eq:repl_running_coupl}
\alpha_a(Q^2) \to \alpha_a(m_W^2)\left(1-\frac{\alpha_a(m_W^2) b^{(1)}_a}{2\pi} L(Q,m_W^2)\right)
\end{equation}
in the Born subamplitude of Eqs.~\eqref{eq:resum_master}--\eqref{eq:resum_allorder}.

Combining the LL and angular independent NLL contributions, the complete angular-independent NLL-accurate resummed correction for each line yields at the \textit{amplitude-squared level} 
\begin{equation}\label{eq:resum_master_squared}
d\sigma^{\rm RG} = \prod_{j=1}^{n} \exp{\left(-\delta^{\rm a.i.}_j(s,M^2)\right)} \cdot d\sigma^{\rm 0} \,,
\end{equation}
where we differentiate external  fermions $F$ and transverse vector bosons $V_{\rm T}$ to yield  \footnote{Note that in $\delta_{V_{\rm T},j}^{\rm a.i.}$ no explicit $U(1)$ contribution  is present since all gauge bosons, both in the broken and unbroken phase, have vanishing hypercharge.} 
\begin{align}\label{eq:full_NLL_resum}
\delta_{F,j}^{\rm a.i.}(s,M^2) = &  \frac{\alpha}{4\pi} C_j \left[ L^2 -3 L \right] \nonumber \\
&- \frac{\alpha^2}{48\pi^2}\left[T_j (T_j+1) \frac{b_2^{(1)}}{s_{\text{w}}^2} +\left( \frac{Y_j}{2}\right)^2 \frac{b_1^{(1)}}{c_{\text{w}}^2} \right] L^3  \,, \nonumber\\
\delta_{V_T,j}^{\rm a.i.}(s,M^2) = &  \frac{\alpha}{4\pi} \left[ \frac{T_j (T_j+1)}{s_{\text{w}}^2} L^2 \right.\nonumber \\
&\left.-\left(\delta_{j,W}b_2^{(1)}+\delta_{j,B} b_1^{(1)}\right)L\right] \nonumber \\
&- \frac{\alpha^2}{48\pi^2}\left[T_j (T_j+1) \frac{b_2^{(1)}}{s_{\text{w}}^2} \right] L^3  \,.
\end{align}

Again, the replacement of Eq.~\eqref{eq:repl_running_coupl} must be appropriately taken into account also in Eq.~\eqref{eq:resum_master_squared} if the renormalisation is performed at scales different from the hard scale.

\subsubsection*{Angular-dependent NLL resummation}

In addition to the angular-independent contributions discussed above, the all-order resummation of angular- dependent (\ad) NLL corrections has been addressed in~\cite{Melles:2001dh}. These corrections arise from soft wide-angle emission of virtual gauge bosons exchanged between different external legs and their resummation involves a non-trivial operator structure in the space of amplitudes with different $SU(2)$ quantum numbers. 

Including the angular-dependent soft-collinear logarithms, the resummed amplitude takes the form
\begin{equation}\label{eq:resum_angular}
\mathcal{M}^{\rm RG\, \ad} = \exp\left[\sum_{j=1}^n \sum_{k<j} \sum_a \frac{\alpha_a}{4\pi} \mathbf{T}^a_j \cdot \mathbf{T}^a_k \, l_{jk}L \right]  \cdot \mathcal{M}_0 \,,
\end{equation}
where $\mathbf{T}^a_j$ denotes the gauge-group generator in the representation of external particle $j$, and the product $\mathbf{T}^a_j \cdot \mathbf{T}^a_k$ acts as an operator on the colour/isospin space of the Born amplitude. In the case of $SU(2)$, this operator structure is responsible for the mixing between different isospin amplitudes and leads to the appearance of $SU(2)$-flipped Born amplitudes in the correction factors.

In our study, the angular-dependent NLL contributions are instead generated via a
diagrammatic approach in the broken EW theory, as discussed in detail in
Sections \ref{sec:oneloop}-\ref{sec:twoloop}. This approach offers several advantages with respect to
a direct expansion of Eq.~\eqref{eq:resum_angular}. First, by working directly in
the broken theory, effects of EW symmetry breaking are automatically accounted for
at the diagrammatic level, and the matching between the symmetric-theory generators
$\mathbf{T}^a_j$ and the physical couplings of the mass-eigenstate fields $A, Z,
W^\pm$ is built in from the outset, without requiring an explicit rotation to the
mass-eigenstate basis. Second, the diagrammatic approach naturally produces results
expressed in terms of the same EW coupling structures and $\text{SU}(2)$-flipped
Born matrix elements that already appear in the one-loop implementation of
Ref.~\cite{Lindert:2023fcu}, allowing for a seamless and unified extension to two
loops within the existing \OpenLoops framework. Third, the explicit computation in
the broken theory provides a highly non-trivial cross-check of the resummed result
\eqref{eq:resum_angular}: the agreement between the two approaches, which requires
extensive cancellations of symmetry- breaking effects among individual diagrams as
demonstrated in \linebreak Refs.~\cite{Denner:2003wi,Denner:2006jr}, confirms the validity of
both the diagrammatic calculation and the IREE resummation at two loops. This agreement
between the diagrammatic broken and resummed unbroken theories has been further validated
by us including for $2 \to 3$ processes as discussed in Section~\ref{sec:validation}.

%%%%%%%%%%%%%%%%%%%%%%%%%%%%%%%%%%%%%%%%%%%%%%%%%%%%%%%%
\subsection{Diagrammatic EW Sudakov corrections at one-loop} \label{sec:oneloop}
%%%%%%%%%%%%%%%%%%%%%%%%%%%%%%%%%%%%%%%%%%%%%%%%%%%%%%%%

%%%%%%%%%%%%%%%%%%%%%%%%%%%%%%%%%%%%%%%%%%%%%%%%%%%%%%%%
\subsubsection{Soft-collinear logarithms}\label{sec:one_loop_factorisable}
%%%%%%%%%%%%%%%%%%%%%%%%%%%%%%%%%%%%%%%%%%%%%%%%%%%%%%%%

The only soft-collinear diagram at one loop is given by the exchange of a soft-collinear gauge boson $V$ between external legs $j$ and $k$:
 \begin{align} \label{diag:one-loop}
\mathcal{M}^{(1), jk} &=    \vcenter{\hbox{\begin{tikzpicture}
    \begin{feynman}
      \vertex (a) at ( 1, 0);
      \vertex (c) at ( -1.5, 1.5);
      \vertex (t) at ( 2, 1);
      \vertex (r) at ( 2, -1);
      \vertex (k) at ( -1, 1); % {\contour{black}{}};
      \vertex (z) at ( -1, -1); % {\contour{black}{}};
      \vertex (v) at ( -1.3, 0) {\(V\)}; 
      \vertex (d) at ( -1.5, -1.5);
      \vertex (jname) at ( -1.15, 1.5) {\(j\)}; 
      \vertex (kname) at ( -1.15, -1.4) {\(k\)};       
      \vertex (jpname) at ( -0.45, 1.0) {\(j^\prime\)}; 
      \vertex (kpname) at ( -0.45, -1.0) {\(k^\prime\)};       
        \vertex[draw,circle,minimum size=0.75cm] (q) at ( 0, 0) {\contour{black}{}};
      \diagram* {
    (q)--[plain] (z),
       (z) -- [plain] (d),
        (q) -- [plain] (k),
       (k) -- [plain]  (c),
       (z) -- [photon]  (k)
         };
\end{feynman}
  \end{tikzpicture}}}
\end{align}

In the eikonal approximation, this yields the correction~\cite{Denner:2000jv}
\begin{equation} \label{DLiniz}
\begin{aligned}
\hat\delta^{\text{SC}, V}_{jk}
\mathcal{M}
_0 \overset{\text{LA}}{=} \mathcal{M}_{0}
\frac{1}{16 \pi^2}
\sum_{j^{\prime}, k^{\prime}}
 & 4p_j p_k \, I_{jj^\prime}^{V} I_{kk^\prime}^{{V}} C_0^{\text{LA}} 
 \,,
\end{aligned}
\end{equation}
where $I_{jj^\prime}^{V}$ are the scalar parts of the tree-level EW vertices $\phiphantJ$-$\varphi_{{j}^{\prime}}$-$V$, implicitly encoding a factor $e$ and sign factors for charged $W$-boson exchange. In the case of a soft $W^{\pm}$ exchange, the external states $\phiphantJ$, $\phiphantK$ are replaced by their $SU(2)$-flipped partners $\varphi_{{j}^{\prime}}$, $\varphi_{{k}^{\prime}}$, so that the matrix element on the rhs of Eq.~\eqref{DLiniz} differs from the Born \footnote{Unless strictly necessary, for improved readability in the following we drop prime indices in diagrams and couplings. Sums over such indices, as in Eq. \eqref{DLiniz}, will be implicitly understood.}.

The scalar three-point function in the high-energy limit reads~\cite{Roth:1996pd}
\begin{equation} \label{C0imaginary} 
C_0^{\text{LA}} =\frac{1}{2 r_{jk}}
  \left[ \log ^{2}\frac{|r_{jk}|}{m_{V}^{2}} - 2 i\pi \Theta(r_{jk})\log \frac{|r_{jk}|}{m_{V}^2}   \right] ,
\end{equation}
and the total one-loop soft-collinear correction sums over all gauge-boson exchanges between all pairs of external legs,
\begin{equation} \label{DLtotal}
\hat\delta^{\text{DL}}\mathcal{M}_0\overset{\text{LA}}{=}
\sum_{j=1}^{n} \sum_{k<j} \sum_{V=A,Z,W^{\pm}} 
\hat\delta_{jk}^{\mathrm{DL}, V}
\mathcal{M}_0\,.
\end{equation}

For the separation into angular-independent and angular-dependent contributions, the log-squared in Eq.~\eqref{C0imaginary} can be decomposed as
\begin{equation} \label{log}
\begin{aligned}
    \log ^{2}\left(\frac{\left|r_{jk}\right|}{m_V^{2}}\right) =&
    \underbrace{\log ^{2}\left(\frac{Q^2}{m_V^{2}}\right)}_{\text{LSC}}+2 \underbrace{\log\left(\frac{Q^2}{m_V^{2}}\right) \log \left(\frac{\left|r_{jk}\right|}{Q^2}\right)}_{\text{SSC}}\\
   & +\underbrace{\log ^{2}\left(\frac{\left|r_{jk}\right|}{Q^2}\right)}_{\text{S-SSC}} = L^2+2\,l_{jk} \, L+l_{jk}^2 \,,
\end{aligned}
\end{equation}
isolating the (angular-independent) leading soft-collinear (LSC) term which represents the LL contribution, the angular-dependent sub-leading soft-collinear (SSC) term which contributes at NLL, and the sub-sub-leading soft-collinear (\SSSC) term. The latter does not contribute in the strict LA limit~\eqref{la} and
thus only yields a NNLL contribution. Therefore, this term has been omitted in Ref.~\cite{Denner:2000jv}, however, it can become numerically relevant when $r_{jk} \gg r_{j'k'}$, as noted in Ref.~\cite{Pagani:2021vyk}. Our implementation in \OpenLoops  computes the LSC, SSC and \SSSC terms individually or their sum via direct application of Eq.~\eqref{DLtotal}.

\subsubsection{Angular-independent NLL}\label{subsec:no_coll_oneloop}

Single-logarithmic (SL) angular-independent one-loop corrections contributing at NLL arise from three sources: the UV regime via renormalisation of dimensionless parameters (PR), wavefunction renormalisation constants (WFRCs), and collinear emission from external states (COLL). The WFRC and COLL contributions factorise with respect to the Born amplitude and together form a gauge-invariant collinear (C) correction associated to external states. The total one-loop SL correction thus decomposes as
\begin{equation}
\label{slsplitting}
\hat\delta^{\mathrm{SL}}= \hat\delta^{\mathrm{PR}}+ \hat\delta^{\mathrm{COLL}}+
\hat\delta^{\mathrm{WFRC}}.
\end{equation}
While the individual contributions are scheme- and scale-dependent -- in particular $\hat\delta^{\mathrm{PR}}$ depends on the renormalisation scale $\mur$ at which the EW couplings are renormalised -- their sum $\hat\delta^{\mathrm{SL}}$ is scheme independent.

The explicit WFRC correction for external fermions and transverse vector bosons, regularised in MR with $\lambda=m_j=M_W$, where $m_j$ and $\lambda$ are respectively the fictitious fermion and photon masses employed for the regularisation of infrared singular photon/QED contributions within MR, is given by\footnote{The full expressions in Eq. \eqref{eq:onenloop_wfr}  contain additional logarithmic contributions of the type $\log(\lambda/m_W)$ and, for external fermions, $\log(m_j/m_W)$ (see Eq. 5.29 and 5.52 in \cite{Pozzorini:2001rs}). However, setting $\lambda=m_j=m_W$ no such QED/photon logarithms contribute.}
\begin{align}\label{eq:onenloop_wfr}
\hat\delta^{\mathrm{WFRC}}_{F,j} &= - \frac{\alpha}{4\pi} \frac{C_j}{2}L \nonumber\\
\hat\delta^{\mathrm{WFRC}}_{V_T, j} &\equiv \sum_k \hat\delta^{\mathrm{WFRC}}_{V_j V_k}  \nonumber\\
&=\sum_k  \frac{\alpha}{4\pi} \left[\frac{b^{(1)}_{V_j V_k}}{2} -C_{V_j V_k} +\frac{b^{(1)}_{AZ}}{2} E_{V_j V_k}  \right]L \,,
\end{align}
where $E$ is an antisymmetric matrix whose components are all vanishing except
for $E_{AZ} = -E_{ZA} = 1$ while $b^{(1)}_{V_j V_k}$ are the one-loop coefficients of the $\beta$-function given by
\begin{align}\label{eq:oneloop_coeff_beta}
&b_{AA} = -\frac{11}{3}, \quad \quad \quad \quad  b_{AZ}=-\frac{19+22s^2_{\text{w}}}{6s_{\text{w}}c_{\text{w}}}  \nonumber \\
&b_{ZZ} = \frac{19-38s^2_{\text{w}} - 22s^2_{\text{w}}}{6s^2_{\text{w}}c^2_{\text{w}}} , \quad \quad b_{WW} = \frac{19}{6s^2_{\text{w}}}.    
\end{align}
In Ref.~\cite{Lindert:2023fcu}, these WFRC corrections were constructed in terms of standard one-loop self-energy corrections in LA. 

In Ref. \cite{Denner:2000jv,Pozzorini:2001rs} it has been explicitly demonstrated that processes involving an arbitrary number $n_{V_{\rm{T}}}$ of external transversely polarised gauge bosons of the type
\begin{equation}\label{eq:proc_vectors}
f^\kappa_{j,\sigma} \bar f^\kappa_{j^\prime,\sigma^\prime} \to V^{a_1}_{\text{T}} \ldots V^{a_n}_{\text{T}},
\end{equation}
feature an exact cancellation of the one-loop coefficients of the $\beta$-function contained in Eq.~\eqref{eq:onenloop_wfr} with the corresponding PR contribution $ \hat\delta^{\mathrm{PR}}$ when this latter is computed within the on-shell renormalisation scheme. This cancellation between the $\beta$-terms of the vector bosons' WFRC correction and corresponding PR contribution of the related $V\bar f f$ coupling in the Born subamplitude, is a consequence of Ward identities and, as such, is exact at each order in perturbation theory.

The decomposition Eq.~\eqref{slsplitting} is directly relevant at two loops: as will be discussed in Section~\ref{sec:twoloop}, the two-loop NLL corrections receive contributions from the product of one-loop counterterms with the one-loop LL amplitude, where the PR and WFRC terms from Eq.~\eqref{slsplitting} enter as building blocks. 

A non-trivial consistency check is provided by the agreement of $\hat\delta^{\mathrm{SL}}$ with the $\mathcal{O}(\alpha)$ expansion of the all-order resummed angular-independent NLL corrections obtained via the IREE~\cite{Fadin:1999bq,Melles:2000gw,Melles:2001mr} in the symmetric $SU(2)\otimes U(1)$ theory, as discussed in Section~\ref{sec:allorders}. We have validated this consistency for several example processes.

%%%%%%%%%%%%%%%%%%%%%%%%%%%%%%%%%%%%%%%%%%%%%%%%%%%%%%%%
\subsection{Diagrammatic EW Sudakov corrections at two-loop} \label{sec:twoloop}
%%%%%%%%%%%%%%%%%%%%%%%%%%%%%%%%%%%%%%%%%%%%%%%%%%%%%%%%

%%%%%%%%%%%%%%%%%%%%%%%%%%%%%%%%%%%%%%%%%%%%%%%%%%%%%%%%
\subsubsection{Soft-collinear logarithms}\label{sec:two_loop_factorisable}
%%%%%%%%%%%%%%%%%%%%%%%%%%%%%%%%%%%%%%%%%%%%%%%%%%%%%%%%

At the two-loop level, NLL accurate terms are extracted from diagrams in which
two vector bosons $V_1$ and $V_2$ are exchanged between external legs, in the
soft and/or collinear regions of the corresponding loop momenta. The extraction
of the logarithmic contributions from these two-loop integrals in the eikonal
approximation has been performed analytically in Ref.~\cite{Denner:2003wi} using sector decomposition 
techniques.

In total there are fourteen distinct topologies, classified according to how many
external legs participate in the exchange of $V_1$ and $V_2$: eleven involve two
external legs, two correspond to exchanges between three external legs, and one
involves four external lines. The full list of these topologies is given in
App.~\ref{app:factorisable_diags}. However, as shown explicitly in
Ref.~\cite{Denner:2006jr}, EW Ward Identities induce substantial cancellations
when all diagrams \eqref{diag:D1}--\eqref{diag:D14} are summed together. The
result is considerably simplified and, at NLL accuracy, can be expressed
in terms of the one-loop integral \footnote{The function $D_0$ is related to the
sum of \LL, the angular dependent NLL and \COLL at one-loop, see Ref.
\cite{Denner:2006jr}.}
\begin{align}\label{eq:d0}
D_0(m_V; r_{jk} ) \overset{\nll}{=}&-L^2 +2\left(2-l_{jk} \right) L +2 i\pi \theta(r_{jk}) l_{jk}(m_V) ,
\end{align}
and one-loop $\beta-$function coefficients introduced in Eq.~\eqref{eq:oneloop_coeff_beta}. $L = L(Q,m_V^2)$ and $l_{jk} = l(r_{jk},Q)$ are defined in \eqref{eq:logs_def_generic}.
The integral $D_0$ in Eq.~\eqref{eq:d0} is defined for spacelike kinematics and requires analytic continuation 
to the physical region. This is achieved via the standard prescription $r_{jk} \to r_{jk} + i0^+$, 
which generates the imaginary term proportional to $\theta(r_{jk})$, where the Heaviside function $\theta(r_{jk})$ selects the timelike configurations $r_{jk}>0$, for which $\log(r_{jk}/Q^2) = l_{jk} + i\pi$, so that the analytic continuation produces a non-vanishing imaginary part.
This imaginary part contributes for $2 \to  n$ processes with $n \ge 3$ and is included throughout in our construction.

Overall, the following three classes of topologies generate all soft-collinear 
two-loop contributions up to NLL:
\begin{itemize}
\item Diagrams involving two external lines $D_1$--$D_{11}$ 
(see \eqref{diag:D1}--\eqref{diag:D6}):
\begin{align}\label{eq:simplified22}
\calM^{(2),jk}_{2-\text{legs}}  \overset{\text{NLL}}{=} & \sum_{p=1}^{11} \calM^{(2),jk}_p  \sim\calM_0 \sum_{V_1,V_2=A,W^\pm,Z}  \nonumber  \\ 
&\left[\frac{1}{2} I_j^{\bar V_1} I_j^{\bar V_2} I_k^{V_1} I_k^{V_2} D_0 (m_{V_1}; r_{jk})  D_0 (m_{V_2}; r_{jk})   \right. \nonumber \\ 
&\left. +\frac{2}{3}I_j^{V_1} I_k^{\bar V_2} b_{V_1 V_2}^{(1)}L^3 \right].
\end{align}
Here the symbol ``$\sim$'' indicates that \eqref{eq:simplified22} receives
additional contributions which cancel exactly against corresponding terms in
Eq.~\eqref{eq:simplified211} below and are therefore omitted. 

\item Diagrams related to three external lines $D_{12}$--$D_{13}$  (see \eqref{diag:D12}--\eqref{diag:D13}):

\begin{align}\label{eq:simplified211}
\calM^{(2),jkg}_{3-\text{legs}}  \overset{\text{NLL}}{=} & \sum_{p=12}^{13} \calM^{(2),jkg}_p \sim\calM_0\sum_{\pi(j,k,g)}\sum_{V_1,V_2=A,W^\pm,Z} \nonumber \\ 
& \left[ \frac{1}{2} I_j^{\bar V_1} I_j^{\bar V_2} I_k^{V_2} I_g^{V_1} D_0 (m_{V_1}; r_{jk})  D_0 (m_{V_2}; r_{jg})  \right]\,,
\end{align}
where the sum runs over all six permutations $\pi(j,k,g)$ of external lines $j,k,g$. Here, terms that cancel against Eq.~\eqref{eq:simplified22} have been omitted.

\item Diagrams related to four external lines $D_{14}$ (Eq. \eqref{diag:D14}):
\begin{align}\label{eq:simplified1111}
\calM^{(2),jkgh}_{4-\text{legs}}  \overset{\text{NLL}}{=} &\calM_0\sum_{V_1,V_2=A,W^\pm,Z}  I_j^{\bar V_1} I_k^{ V_1} I_g^{\bar V_2} I_h^{V_2}  \nonumber  \\ 
&D_0 (m_{V_1}; r_{jk})  D_0 (m_{V_2}; r_{gh}) \,.
\end{align}
\end{itemize} 

By summing Eqs. \eqref{eq:simplified22}, \eqref{eq:simplified211} and \eqref{eq:simplified1111} we can define the total two-loop LL and NLL angular dependent (NLL a.d.) corrections as 

\begin{align}\label{eq:ll_2loop_total}
\hat\delta^{(2)}_{\text{LL}} \,\mathcal{M}_0 =& \sum_{j=1}^n\sum_{k<j}^n \left[\calM^{(2),jk}_{2-\text{legs}} + \sum_{g<k}^n \calM^{(2),jkg}_{3-\text{legs}} \right.     \nonumber \\ 
&\left. \left.+\sum_{g<k}^n\sum_{h<g}^n\calM^{(2),jkgh}_{4-\text{legs}} \right]\right|_{L^4} \quad \phantom{a,b\in \{j,k,g,h\}}\\
\label{eq:nll_2loop_total}
\hat\delta^{(2)}_{\text{NLL a.d.}} \,\mathcal{M}_0 =& \sum_{j=1}^n\sum_{k<j}^n  \left[\calM^{(2),jk}_{2-\text{legs}} + \sum_{g<k}^n \calM^{(2),jkg}_{3-\text{legs}}  \right.   \nonumber \\ 
&\left.\left.+\sum_{g<k}^n\sum_{h<g}^n\calM^{(2),jkgh}_{4-\text{legs}} \right]\right|_{l_{ab}L^3}, \end{align}
where $a,b\in \{j,k,g,h\}$.

The sum of Eqs. \eqref{eq:simplified22}, \eqref{eq:simplified211} and \eqref{eq:simplified1111} also yields NLL terms of collinear nature, originating from configurations where one of the gauge bosons is soft-collinear while the other is collinear to an external legs. This contribution can be isolated as 
\begin{align}\label{eq:ll_2loop_total}
\hat\delta^{(2)}_{\text{coll}} \,\mathcal{M}_0 =& \sum_{j=1}^n\sum_{k<j}^n \left[\calM^{(2),jk}_{2-\text{legs}} + \sum_{g<k}^n \calM^{(2),jkg}_{3-\text{legs}} \right.     \nonumber \\ 
&\left. \left.+\sum_{g<k}^n\sum_{h<g}^n\calM^{(2),jkgh}_{4-\text{legs}} \right]\right|_{L^3} \quad \phantom{a,b\in \{j,k,g,h\}}
 \end{align}
and feeds into the total NLL angular independent (NLL a.i.) correction defined below in Eq.~\eqref{eq:ai_nll_2loop}

%%%%%%%%%%%%%%%%%%%%%%%%%%%%%%%%%%%%%%%%%%%%%%%%%%%%%%%%
\subsubsection{Angular-independent NLL}\label{sec:two_loop_ai_nll}
%%%%%%%%%%%%%%%%%%%%%%%%%%%%%%%%%%%%%%%%%%%%%%%%%%%%%%%%

At two loops, additional NLL angular-independent contributions arise from the combination of 
the one-loop LL amplitude with one-loop counterterms and WFRC contributions, and from the one-loop renormalisation of the one-loop insertion itself. 
Mass renormalisation generates non-suppressed logarithmic terms only 
through the insertion of one-loop counterterms into the one-loop logarithmic corrections; 
these are of NNLL order and can therefore be neglected at NLL~\cite{Pozzorini:2004rm}. 
Likewise, the genuinely two-loop counterterms associated with external-field wave-function 
and coupling renormalisation are of NNLL order.

The first two-loop NLL PR contribution is obtained by inserting 
the one-loop coupling counterterms into the one-loop LL amplitude,
\begin{equation}\label{eq:pr_2loop}
\hat\delta^{(2)}_{\text{PR1}} \,\mathcal{M}_0  \overset{\text{NLL}}{=}  \hat\delta^{(1)}_{\mathrm{PR}} \cdot \hat\delta^{(1)}_{\mathrm{LL}} \,\mathcal{M}_0 \,.
\end{equation}
This accounts for the renormalisation of the Born couplings. As at one-loop we construct this contribution from the standard one-loop on-shell renormalisation implemented in 
\OpenLoops~\cite{Buccioni:2019sur}, restricted to the LA.\footnote{In Refs.~\cite{Denner:2006jr,Denner:2008yn} this contribution vanishes as consequence of the choice $\mur=Q$ for the renormalisation of Born couplings.}

The second NLL PR contributions originates from the one-loop renormalisation of the 
couplings associated with the genuine one-loop insertion itself. This contribution can be recast as~\cite{Denner:2006jr}
\begin{align}\label{eq:pr_1loopins}
\hat\delta^{(2)}_{\text{PR2}}  & \overset{\text{NLL}}{=} \sum_{j=1}^n \sum_{V_1,V_2=A,W^\pm,Z} \nonumber \\
-\frac{1}{2}&\left[  b_{V_1 V_2}^{(1)} I_j^{V_1}  I_j^{\bar V_2}    \right] 
D_0(m_V; -Q^2)\Big|_{\text{LL}}\, 
L(Q,\mur^2)\,\mathcal{M}_0.
\end{align}

The entire PR contribution is then obtained as
\begin{align}
 \hat\delta^{(2)}_{\mathrm{PR}} =  \hat\delta^{(2)}_{\mathrm{PR1}}	+ \hat\delta^{(2)}_{\mathrm{PR2}}
\end{align}
and depends on the renormalisation scale 
$\mur$.

As for the PR the two-loop NLL WFRC contribution arises from inserting 
the one-loop correction $\hat\delta^{(1)}_{\mathrm{WFRC}}$, Eq.~\eqref{slsplitting}, 
into the one-loop LL amplitude,
\begin{equation}\label{eq:wfr_2loop}
\hat\delta^{(2)}_{\text{WFRC}} \,\mathcal{M}_0  \overset{\text{NLL}}{=}  \hat{\delta}^{(1)}_{\mathrm{WFRC}} \cdot \hat\delta^{(1)}_{\mathrm{LL}} \,\mathcal{M}_0 \,,
\end{equation}
where $\hat{\delta}^{(1)}_{\mathrm{WFRC}}$ collects the one-loop wave-function 
renormalisation logarithms for each external leg.

For processes with external transversely polarised gauge bosons, the one-loop $\beta$-function coefficients entering the vector 
bosons' WFRC in Eq.~\eqref{eq:onenloop_wfr} are cancelled exactly by $\hat\delta^{(1)}_{\mathrm{PR}}$ 
from renormalisation of the $V\bar{f}f$ coupling in the Born subamplitude, when employing 
on-shell renormalisation. This cancellation holds to all orders in perturbation theory. To ensure this cancellation at two-loop we define $\hat{\delta}^{(1)}_{\mathrm{WFRC}}$ entering in Eq. \eqref{eq:wfr_2loop} as
\begin{equation}
\hat \delta^{(1)}_{\mathrm{WFRC}}  \overset{\text{NLL}}{=}  \sum_{i=1}^{n_F}  \hat\delta^{\mathrm{WFRC}}_{F,j}  + \sum_{i=1}^{n_{V_{\rm{T}}}}  \hat{\tilde{\delta}}^{\mathrm{WFRC}}_{V_j}\,
\end{equation}
summing over $n_F$ external fermions and $n_{V_{\text{T}}}$ external transverse vector-bosons. Here the vector-boson WFRC is modified as 
\begin{equation}
\hat{\tilde{\delta}}^{\mathrm{WFRC}}_{V_j}  \overset{\text{NLL}}{=}   \sum_k  \frac{\alpha}{4\pi} \left[ -C_{V_j V_k}  \right]L\,,            
\end{equation}
i.e.\ using the one-loop WFRC of Eq.~\eqref{eq:onenloop_wfr} with all $b_{V_j V_k}$ terms 
removed. Simultaneously, the PR contribution from renormalisation of the $V_j\bar{f}f$ 
coupling in the Born subamplitude is switched off. 
Thus, we assume the cancellation of 
$\beta$-dependent terms to hold to all orders, as predicted by the IREE resummation in 
the symmetric $SU(2)\otimes U(1)$ theory~\cite{Melles:2000gw,Melles:2001mr}.
Although the IREE argument is formulated in the unbroken symmetric phase, the 
cancellation is expected to persist in the broken phase at NLL accuracy, since 
mass-suppressed corrections entering through the symmetry-breaking scale $m_W/Q$ 
are beyond NLL and can therefore be neglected. 
For general processes involving both external vector bosons and fermions
we have to distinguish the $n_{V_{\text{T}}}$ Born couplings for which we assume exact cancellation of  $\beta$-dependent terms in the on-shell scheme.

The complete two-loop angular-independent NLL correction is constructed as,
\begin{equation}\label{eq:ai_nll_2loop}
\hat\delta^{(2)}_{\mathrm{NLL,\, a.i.}} =\hat\delta^{(2)}_{\text{coll}}  +  \hat\delta^{(2)}_{\text{WFRC}} + \hat\delta^{(2)}_{\text{PR}} \,.
\end{equation}

This construction of angular-independent NLL corrections can be validated against the $\mathcal{O}(\alpha^2)$ expansion of Eq.~\eqref{eq:full_NLL_resum}. In fact, this allows to check the entire $\beta$-function--dependent terms originating from the two-loop soft-collinear  diagrams 
\eqref{eq:simplified22} and the angular-independent NLL contribution.

Finally, we define the total two-loop NLL correction as sum of the NLL angular-dependent, Eq.~\eqref{eq:nll_2loop_total}, and NLL angular-independent, Eq.~\eqref{eq:ai_nll_2loop}, contributions as
\begin{equation}\label{eq:nll_2loop_total}
\hat\delta^{(2)}_{\mathrm{NLL}} = \hat\delta^{(2)}_{\mathrm{NLL,\, a.d.}}  + \hat\delta^{(2)}_{\mathrm{NLL,\, a.i.}}  \,.
\end{equation}

%%%%%%%%%%%%%%%%%%%%%%%%%%%%%%%%%%%%%%%%%%%%%%%%%%%%%%%%
\section{Implementation in OpenLoops} \label{sec:implementation}
%%%%%%%%%%%%%%%%%%%%%%%%%%%%%%%%%%%%%%%%%%%%%%%%%%%%%%%%

In the following we concisely review the structure of the implementation of one-loop EW logarithms presented in \cite{Lindert:2023fcu, Mai:2024fsk},  its extension  to the two-loop level, and validation against analytical results.

%%%%%%%%%%%%%%%%%%%%%%%%%%%%%%%%%%%%%%%%%%%%%%%%%%%%%%%%
\subsection{One-loop implementation}
%%%%%%%%%%%%%%%%%%%%%%%%%%%%%%%%%%%%%%%%%%%%%%%%%%%%%%%%

At one loop, the angular-dependent soft-collinear contribution,
Eq.~\eqref{DLiniz}, requires a double sum over all pairs of
external legs and all soft EW vector bosons $V=A,Z,W^\pm$, involving
the evaluation of LO Born matrix elements as well as
$\text{SU}(2)$-flipped tree amplitudes for $W^\pm$-boson exchange.
Similarly, the LL and angular-independent NLL contributions involve
single sums over external legs with, in general,
$\text{SU}(2)$-flipped Born matrix elements. We implement all such
sums via effective helicity-dependent EW Sudakov external-leg
pseudo-counterterms, which can be generated automatically for any
process in \OpenLoops.

The key observation is that, in the logarithmic approximation, the
virtual soft-$V$ propagator together with its adjacent internal
propagators can be replaced by a correlator of two external-leg
pseudo-counterterms. Taking the Drell--Yan process
$q\bar{q}\to\ell^+\ell^-$ as an illustrative example, this
replacement reads
\begin{equation}
    \vcenter{\hbox{\begin{tikzpicture}
     \begin{feynman}
      \vertex (a) at ( 1, 0);
      \vertex  (e) at ( 0, 0);
       \vertex (k) at ( -0.5, 0.5);
       \vertex (o) at ( -0.7, 0.1);
       \vertex (i) at ( -0.7, -0.1);
      \vertex (z) at ( -0.5, -0.5);
      \vertex (c) at ( -1, 1);
      \vertex (d) at ( -1, -1);
      \vertex (f) at (2, 1);
      \vertex (g) at (2, -1);
      \diagram* {
        (a) -- [photon,edge label'=\( V^\prime\)]
         (e),
         (e) -- [plain, edge label=\(\bar q^{\prime}\)]
        (z),
          (z) -- [plain, edge label=\(\bar q\)]
         (d),
             (e) -- [plain, edge label' =\(q^{\prime}\)]
         (k),
          (k) -- [plain, edge label' =\(q\)]
         (c),
         (a) -- [plain, edge label=\( \ell^+\)]
         (f),
             (a) -- [plain, edge label'=\( \ell^-\)]
         (g),
         (z) -- [photon,edge label=\(V\)]
         (i),
             (k) -- [photon,edge label'=\(V\)]
         (o),
         };
    \end{feynman}
  \end{tikzpicture}}}
     \hspace{0.1 cm} \overset{\text{CT}}{\longrightarrow}
     \vcenter{\hbox{\begin{tikzpicture}
     \begin{feynman}
      \vertex (a) at ( 1, 0);
      \vertex  (e) at ( 0, 0);
      \vertex (c) at ( -1, 1);
      \vertex (t) at ( 2, 1);
      \vertex (r) at ( 2, -1);
      \vertex[dot] (k) at ( -0.5, 0.5) {\contour{black}{}};
      \vertex[dot] (z) at ( -0.5, -0.5) {\contour{black}{}};
     \vertex (u) at (-0.8, -0.5) {\(V\)};
       \vertex (u) at (-0.8, 0.45) {\(V\)};
      \vertex (d) at ( -1, -1);
      \diagram* {
        (a) -- [photon,edge label'=\( V ^\prime\)]
         (e),
           (e) -- [plain,  insertion={[size=2 pt, style=thick]0.5}, edge label=\(\bar q^{\prime}\)]
         (z),       
          (z) -- [plain, edge label=\(\bar q\)]
         (d),
             (e) -- [plain,  insertion={[size=2 pt, style=thick]0.5 }, edge label' =\(q^{\prime}\)]
         (k),
          (k) -- [plain, edge label'=\(q\)]
         (c),
         (a) -- [plain,  edge label=\( \ell^+\)]
         (t),
         (a) -- [plain,  edge label'=\( \ell^-\)]
         (r)
         };
    \end{feynman}
  \end{tikzpicture}}}
   \end{equation}
where the crosses on the fermion lines indicate removed propagators,
and effects due to mass differences $m_q \neq m_{q'}$ are
mass-suppressed. To automate this procedure in \OpenLoops, we define
helicity-dependent effective two-point vertex rules,
\begin{equation} \label{eq:effectiveCTrule_fermions}
 \vcenter{\hbox{\begin{tikzpicture}
     \begin{feynman}
      \vertex (a) at ( 1, 0);
      \vertex  (e) at ( 0, 0);
      \vertex (c) at ( -2, 0);
      \vertex (u) at (-1, 0.75);
      \vertex (k) at ( -1, 0);
      \diagram* {
             (e) -- [plain, edge label=\(f^{\prime \kappa}\)]
         (k),
          (k) -- [plain, edge label=\(f^{\kappa\phantom{\prime}}\)]
         (c),
         (k) -- [photon, edge label=\(V\)]
         (u)
         };
    \end{feynman} $\hspace{0.5 cm} \longrightarrow \hspace{0.5 cm}$
  \end{tikzpicture}
 $\hspace{0.5 cm} \hspace{0.8 cm}$
 \begin{tikzpicture}
     \begin{feynman}
      \vertex (a) at ( 1, 0);
      \vertex  (e) at ( 0, 0);
      \vertex (c) at ( -2, 0);
      \vertex[dot] (k) at ( -1, 0) {\contour{black}{}};
       \vertex (u) at (-1, 0.3) {\(V\)};
      \diagram* {
             (e) -- [plain,  insertion={[size=2 pt, style=thick]0.5}, edge label=\(f^{\prime \kappa}\)]
         (k),
          (k) -- [plain, edge label=\(f^{\kappa\phantom{\prime}}\)]
         (c),
         };
    \end{feynman}
  \end{tikzpicture}
  }}
    =
  ie \mathcal{I}_{f^{\kappa}}^{V} \,,
\end{equation}
where the factors $\mathcal{I}_{f^{\kappa}}^{V}$ are derived from
the constant parts of the three-point vertices introduced in Eq.~\eqref{DLiniz}, i.e
\begin{equation}
I_{f^{\kappa}}^V=ie\mathcal{I}^V_{f^{\kappa}}, 
\end{equation}
 and listed in Appendix~A of
Ref.~\cite{Lindert:2023fcu} for all relevant EW Sudakov vertices in
the SM. This construction naturally projects onto external-state
helicities and generates the required $\text{SU}(2)$-flipped Born
matrix elements in a streamlined way.

%%%%%%%%%%%%%%%%%%%%%%%%%%%%%%%%%%%%%%%%%%%%%%%%%%%%%%%%
\subsection{Two-loop implementation}
%%%%%%%%%%%%%%%%%%%%%%%%%%%%%%%%%%%%%%%%%%%%%%%%%%%%%%%%

The pseudo-counterterm approach extends straightfor-\linebreak wardly to two
loops. As discussed in Section~\ref{sec:twoloop}, the full set of
two-loop diagrams \eqref{diag:D1}--\eqref{diag:D14} reduces, after
Ward-identity cancellations, to contributions expressible entirely
in terms of products of one-loop $D_0$ functions,
Eqs.~\eqref{eq:simplified22}--\eqref{eq:simplified1111}. These can
be implemented via three topologies of pseudo-counterterm insertions
on external legs, involving four insertions distributed across two,
three, or four distinct external lines:
\begin{itemize}
\item Two external lines:
 \begin{align} \label{diag:22}
\mathcal{M}_{\text{2-legs}}^{(2), jk} =    \vcenter{\hbox{\begin{tikzpicture}
    \begin{feynman}
      \vertex (a) at ( 1, 0);
      \vertex (c) at ( -1.5, 1.5);
      \vertex (t) at ( 2, 1);
      \vertex (r) at ( 2, -1);
      \vertex[dot] (k) at ( -0.7, 0.7) {\contour{black}{}};
      \vertex[dot](z) at ( -0.7, -0.7){\contour{black}{}};
      \vertex[dot] (k1) at ( -1.25, 1.25) {\contour{black}{}};
      \vertex[dot] (z1) at ( -1.25, -1.25) {\contour{black}{}};
      \vertex (jname) at ( -1.65, 1.6) {\(j\)};
      \vertex (kname) at ( -1.65, -1.6) {\(k\)};
      \vertex (d) at ( -1.5, -1.5);
        \vertex[draw,circle,minimum size=0.75cm] (q) at ( 0, 0) {\contour{black}{}};
      \diagram* {
    (q)--[plain,  insertion={[size=2 pt, style=thick]0.4}] (z),
       (z) -- [plain,  insertion={[size=2 pt, style=thick]0.5}] (z1),
         (z1) -- [plain] (d),
             (q)--[plain,  insertion={[size=2 pt, style=thick]0.4}] (k),
       (k) -- [plain,  insertion={[size=2 pt, style=thick]0.5}] (k1),
         (k1) -- [plain] (c) 
                  };
\end{feynman}
  \end{tikzpicture}}}
  \end{align}

\item Three external lines:
 \begin{align} \label{diag:211}
\mathcal{M}_{\text{3-legs}}^{(2), jkg} =     \vcenter{\hbox{\begin{tikzpicture}
    \begin{feynman}
      \vertex (c) at ( -1.5, 1.5);
      \vertex (t) at ( 2, 1);
      \vertex (r) at ( 2, -1);
      \vertex (a) at ( -1.85, 0);
      \vertex[dot] (k) at ( -1, 1) {\contour{black}{}};
      \vertex[dot] (z) at ( -1, -1) {\contour{black}{}};
      \vertex[dot] (z1) at ( -0.9, 0) {\contour{black}{}};
      \vertex[dot] (z2) at ( -1.60, 0) {\contour{black}{}};
      \vertex (d) at ( -1.5, -1.5);
      \vertex (jname) at ( -1.65, 1.6) {\(k\)};
      \vertex (kname) at ( -2.05, 0) {\(j\)};
      \vertex (kname) at ( -1.65, -1.6) {\(g\)};
        \vertex[draw,circle,minimum size=0.75cm] (q) at ( 0, 0) {\contour{black}{}};
      \diagram* {
    (q)--[plain,  insertion={[size=2 pt, style=thick]0.5}]  (z),
       (z) -- [plain] (d),
        (q) -- [plain,  insertion={[size=2 pt, style=thick]0.5}] (k),
       (k) -- [plain]  (c),
       (a) -- [plain,  insertion={[size=2 pt, style=thick]0.41},
               insertion={[size=2 pt, style=thick]0.88}]  (q)
         };
\end{feynman}
  \end{tikzpicture}}}
  \end{align}

\item Four external lines:
 \begin{align}  \label{diag:1111}
\mathcal{M}_{\text{4-legs}}^{(2), jkgh} =        \vcenter{\hbox{\begin{tikzpicture}
    \begin{feynman}
      \vertex (a) at ( 1, 0);
      \vertex (c) at ( -1.5, 1.5);
      \vertex (t) at ( 2, 1);
      \vertex (r) at ( 2, -1);
      \vertex (a) at ( -1.75, 0.65);
      \vertex (b) at ( -1.75, -0.65);
      \vertex[dot] (k) at ( -1, 1) {\contour{black}{}};     %crossed dot, style=thick, minimum size=6pt
      \vertex[dot] (z) at ( -1, -1) {\contour{black}{}};
      \vertex[dot] (z1) at ( -1.2, -0.44) {\contour{black}{}};
      \vertex[dot] (z2) at ( -1.2, 0.44) {\contour{black}{}};
      \vertex (d) at ( -1.5, -1.5);
      \vertex (jname) at ( -1.65, 1.6) {\(j\)};
      \vertex (jname) at ( -1.95, 0.75) {\(k\)};
      \vertex (jname) at ( -1.95, -0.75) {\(g\)};
      \vertex (kname) at ( -1.65, -1.6) {\(h\)};
        \vertex[draw,circle,minimum size=0.75cm] (q) at ( 0, 0) {\contour{black}{}};
      \diagram* {
    (q)--[plain,  insertion={[size=2 pt, style=thick]0.5}]  (z),
       (z) -- [plain]  (d),
        (q) --[plain,  insertion={[size=2 pt, style=thick]0.5}]  (k),
       (k) -- [plain]  (c),
       (a) -- [plain,  insertion={[size=2 pt, style=thick]0.7}]  (q),
       (b) -- [plain,  insertion={[size=2 pt, style=thick]0.7}]  (q),
         };
\end{feynman}
  \end{tikzpicture}}}
  \end{align}
\end{itemize}
Here we sum over all assignments of external legs to the topologies
\eqref{diag:22}--\eqref{diag:1111} and over all soft boson exchanges
$V_{1,2}=A,Z,W^\pm$, according to
Eqs.~\eqref{eq:simplified22}--\eqref{eq:simplified1111}.

The angular-independent NLL contributions, Eqs.~\eqref{eq:pr_2loop}--\eqref{eq:wfr_2loop}, are implemented via additional pseudo-counterterm insertions according to their structure. The first two-loop NLL PR contribution, Eq.~\eqref{eq:pr_2loop}, is implemented via double insertions on individual external legs in order to isolate the LL one-loop structure, supplemented with an additional insertion on EW vertices as performed for the one-loop PR counterterm structure of
Ref.~\cite{Lindert:2023fcu}:

 \begin{align} \label{diag:pr1}
\mathcal{M}_{\text{PR1}}^{(2), jk} =    \vcenter{\hbox{\begin{tikzpicture}
    \begin{feynman}
      \vertex (a) at ( 1, 0);
      \vertex (c) at ( -1.5, 1.5);
      \vertex (t) at ( 2, 1);
      \vertex (r) at ( 2, -1);
      \vertex[dot] (k) at ( -1, 1) {\contour{black}{}};
      \vertex[dot](z) at ( -1, -1){\contour{black}{}};
      \vertex (jname) at ( -1.65, 1.6) {\(j\)};
      \vertex (kname) at ( -1.65, -1.6) {\(k\)};
      \vertex (d) at ( -1.5, -1.5);
      \vertex[dot] (d2) at ( -0.35, 0) {\contour{black}{}};
        \vertex[draw,circle,minimum size=0.75cm] (q) at ( 0, 0) {\contour{black}{}};
      \diagram* {
    (q)--[plain,  insertion={[size=2 pt, style=thick]0.4}] (z),
         (z) -- [plain] (d),
             (q)--[plain,  insertion={[size=2 pt, style=thick]0.4}] (k),
         (k) -- [plain] (c) 
                  };
\end{feynman}
  \end{tikzpicture}}}
  \end{align}

The second two-loop NLL PR contribution, Eq.~\eqref{eq:pr_1loopins}, is implemented via double insertions on each external leg:

 \begin{align} \label{diag:pr2}
\mathcal{M}_{\text{PR2}}^{(2), j} =    \vcenter{\hbox{\begin{tikzpicture}
    \begin{feynman}
      \vertex (c) at ( -1.5, 1.5);
      \vertex (t) at ( 2, 1);
      \vertex (r) at ( 2, -1);
      \vertex (a) at ( -1.85, 0);
      \vertex[dot] (z1) at ( -0.9, 0) {\contour{black}{}};
      \vertex[dot] (z2) at ( -1.60, 0) {\contour{black}{}};
      \vertex (d) at ( -1.5, -1.5);
      \vertex (kname) at ( -2.05, 0) {\(j\)};
        \vertex[draw,circle,minimum size=0.75cm] (q) at ( 0, 0) {\contour{black}{}};
      \diagram* {
       (a) -- [plain,  insertion={[size=2 pt, style=thick]0.41},
               insertion={[size=2 pt, style=thick]0.88}]  (q)
         };
\end{feynman}
  \end{tikzpicture}}}
  \end{align}
  
Finally, the WFRC contribution, Eq. \eqref{eq:wfr_2loop} is implemented via further four pseudo-counterterm insertions on the external legs. In this way we emulate the structure of a self energy insertion multiplying a one-loop LL diagram. To this end we re-use the topology of Eq.~\eqref{diag:211} and add an additional configuration according to
 \begin{align} \label{diag:wfr1}
    &\mathcal{M}_{\text{WFRC}}^{(2), jk} =   \vcenter{\hbox{\begin{tikzpicture}
    \begin{feynman}
      \vertex (a) at ( 1, 0);
      \vertex (c) at ( -1.5, 1.5);
      \vertex (t) at ( 2, 1);
      \vertex (r) at ( 2, -1);
      \vertex[dot] (k) at ( -0.65, 0.65) {\contour{black}{}};
    \vertex[dot] (z2) at ( -1, 1) {\contour{black}{}};
      \vertex[dot] (k1) at ( -1.35, 1.35) {\contour{black}{}};
      \vertex[dot] (z1) at ( -0.75, -0.75) {\contour{black}{}};
     \vertex (v2) at ( -0.5, 1.3);
      \vertex (d) at ( -1.5, -1.5);
     \vertex (jname) at ( -1.65, 1.6) {\(j\)}; 
      \vertex (kname) at ( -1.65, -1.6) {\(k\)}; 
        \vertex[draw,circle,minimum size=0.75cm] (q) at ( 0, 0) {\contour{black}{}};
      \diagram* {
    (q)--[plain] (z),
       (z1) -- [plain] (d),
        (q) -- [plain,  insertion={[size=2 pt, style=thick]0.5}](k),
       (k) -- [plain,  insertion={[size=2 pt, style=thick]0.2}]  (c),
       (z2) --[plain,  insertion={[size=2 pt, style=thick]0.5}] (k1),
        (z1) -- [plain,  insertion={[size=2 pt, style=thick]0.5}]  (q),
         };
\end{feynman}
  \end{tikzpicture}}} \qquad .
\end{align}

%%%%%%%%%%%%%%%%%%%%%%%%%%%%%%%%%%%%%%%%%%%%%%%%%%%%%%%%
\subsection{Validation}\label{sec:validation}
%%%%%%%%%%%%%%%%%%%%%%%%%%%%%%%%%%%%%%%%%%%%%%%%%%%%%%%%
\begin{figure*}[tbh]
\centering
  \includegraphics[width=\setrelwidth\textwidth]{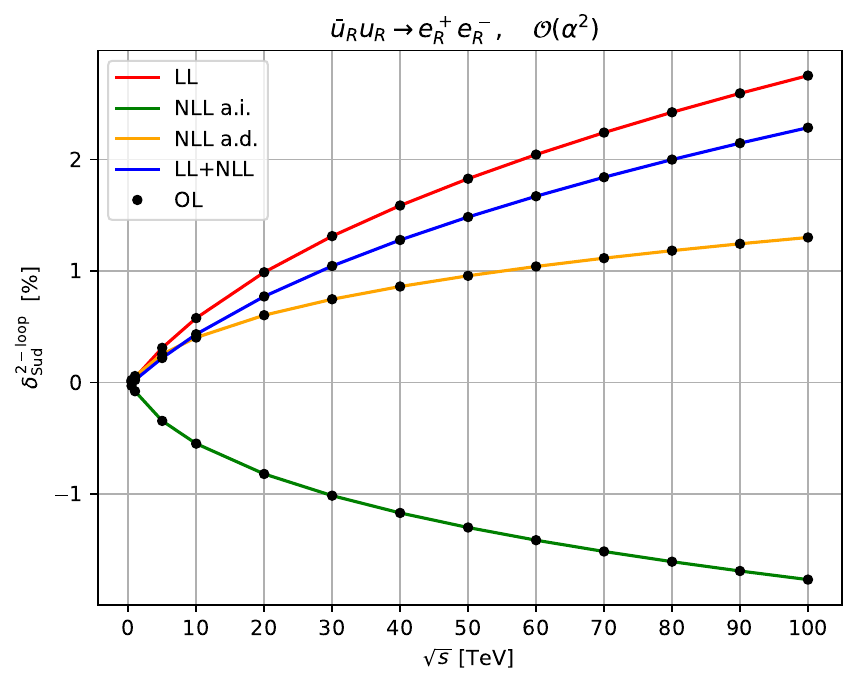}
  \includegraphics[width=\setrelwidth\textwidth]{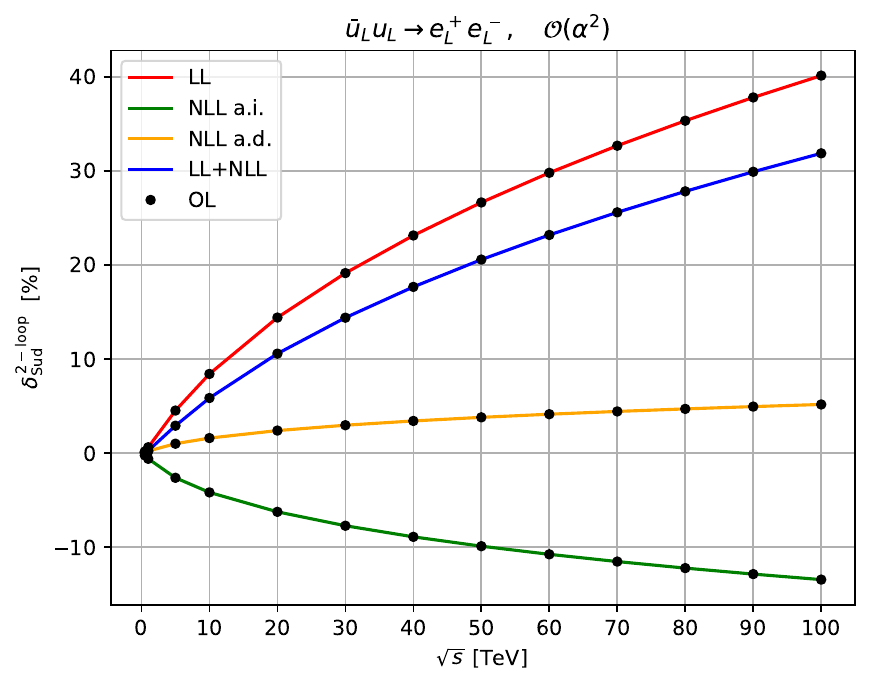}
\caption{Energy scan for $\bar u_R u_R \to e^+_R e^-_R$ and
$\bar u_L u_L \to e^+_L e^-_L$. Shown are the relative two-loop EW
corrections with respect to LO: LL (solid red), NLL a.i.\ (solid
green), NLL a.d.\ (solid orange), and their sum (solid blue). Born
couplings are renormalised at $\muren=\sqrt{s}$; IR divergences are
regularised in MR with a fictitious photon mass $\lambda=m_W$.
Analytical results appear as solid lines; numerical \OpenLoops
predictions as black dots.}
\label{fig:val_qqee}
\end{figure*}
\begin{figure*}[tb]
\centering
  \includegraphics[width=\setrelwidth\textwidth]{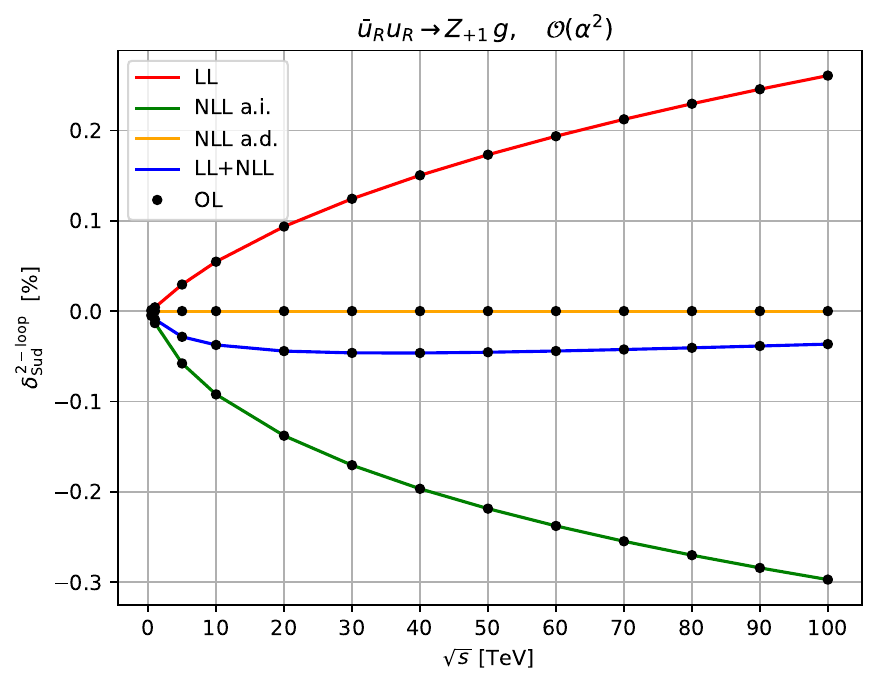}
   \includegraphics[width=\setrelwidth\textwidth]{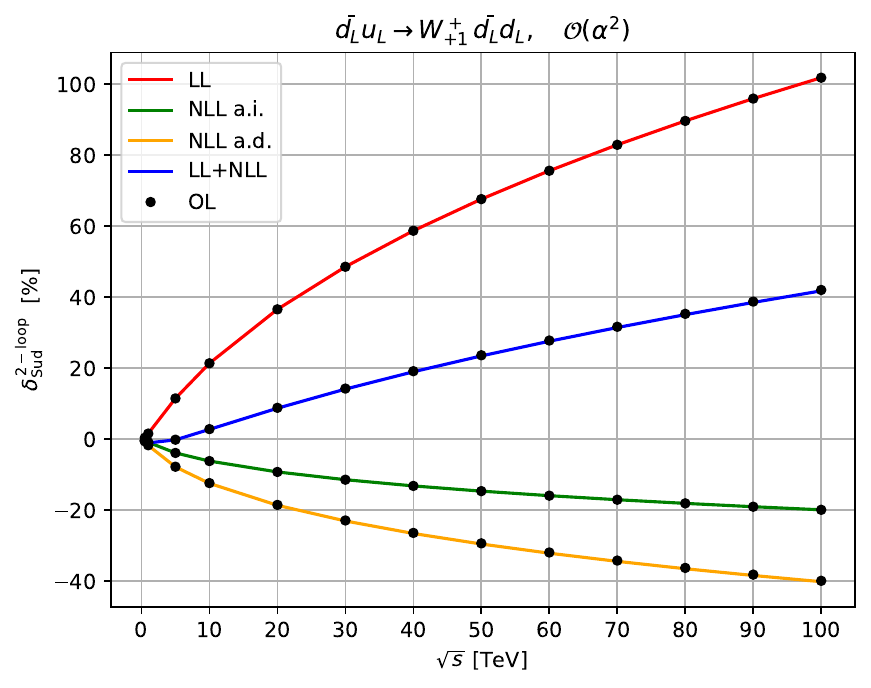}
\caption{Energy scan for $\bar u_R u_R \to Z_{+1}g$ and
%$\bar u_L u_L \to Z_{+1}g$.
$\bar d_L u_L \to W^+_{+1} \bar d_L d_L$. Born
couplings are renormalised at $\muren=m_W$. For the $2\to3$ process we have adopted the following scattering angles: $\theta_{ \bar d}= 58.85^\circ, \,\phi_{ \bar d}= 114,05^\circ, \, \theta_{ d}= 85,13^\circ, \linebreak \phi_{d}= 38,82^\circ, \, \theta_{ W^+}=40,90^\circ, \, \phi_{ W^+}= 99,19^\circ$. Here $\theta_{\varphi}$ and $\phi_{\varphi}$ are, respectively, the polar and azimuthal scattering angles of the momenta $p_\varphi$ of the final state particles with respect to the direction of the incoming particles  with $\varphi \in \{d, \bar d, W^+\}$ in the center of mass frame. Curves as in Fig.~\ref{fig:val_qqee}.}
\label{fig:val_zj_wjj}
\end{figure*}
We validate our implementation by comparing amplitude-squared
predictions against analytical results available in the
literature~\cite{Denner:2006jr, Kuhn:2004em, Kuhn:2005gv,
Kuhn:2007qc}. Specifically, we extract the numerical coefficients of
the individual LL, NLL a.i.\ and NLL a.d.\ logarithmic contributions
through a linear fit in the argument of the respective logarithms and
compare them with the corresponding analytical predictions. These
checks have been performed for all non-mass-suppressed helicity
configurations of the following processes: fermion pair-production
in gluon fusion ($gg \to \bar{u}u$, $gg \to \bar{d}d$);
neutral-current four-fermion processes ($e^+e^- \to \mu^+\mu^-$,
$e^+e^- \to \bar{u}u$, $\bar{\nu}_e\nu_e \to \bar{\nu}_i\nu_i$
with $i\in\{e,\mu\}$); charged-current four-fermion processes
($\bar{\nu}_e e^- \to \bar{u}d$); and $V$+jet production
($\bar{d}u \to W^+g$, $\bar{q}q \to Zg$, $\bar{q}q \to \gamma g$).
Additionally, we have employed the all-order resummation formalism
of Refs.~\cite{Melles:2000gw,Melles:2001mr}, reviewed in
Section~\ref{sec:allorders}, to derive the total two-loop
NLL-accurate corrections to the partonic processes $\bar d u \to W^+ g g$ and
$\bar d u \to W^+ \bar d d$. Excellent agreement is found for all
contributions and processes.

We summarise below the numerical results of these amplitude-level
validations, considering energy scans over
$\sqrt{s} \in [0.5, 100]$~TeV. Scattering angles are kept fixed and set to 
$\theta = 45^\circ$ in $2\to2$ processes, whereas for  $2\to 3$ scatterings we adopt 
scattering angles as defined in the caption of the corresponding plot. 
Polarisation labels follow the convention that
L/R denote left- and right-handed fermions, while $\pm 1$ denote
transverse polarisations of vector bosons.
Figures~\ref{fig:val_qqee}--\ref{fig:val_zj_wjj} show representative
validation plots for $e^+e^- \to \bar{u}u$,
$\bar{q}q \to Zg$, and $\bar d u \to W^+ \bar d d$: analytical
results appear as solid lines (red for LL, green for NLL a.i.,
orange for NLL a.d., blue for their sum), with numerical \OpenLoops
results superimposed as black dots. 
We find the same agreement between analytical results and our \OpenLoops implementation for all other considered processes, chirality configurations and varying scattering angles.

%%%%%%%%%%%%%%%%%%%%%%%%%%%%%%%%%%%%%%%%%%%%%%%%%%%%%%%%
\section{Numerical results} \label{sec:results}

In this section we present numerical results for our implementation of two-loop EW 
corrections in logarithmic approximation in \OpenLoops, covering a representative set 
of processes with massless external fermions and transverse gauge bosons. We investigate 
the impact of two-loop logarithmic corrections relative to their one-loop counterparts in logarithmic approximation and at the full one-loop level.

To isolate the virtual EW corrections, at one-loop infrared divergences are regulated dimensionally 
and the corresponding poles are removed by adding the Catani--Seymour QED 
$\mathbf{I}^{\rm QED}$-operator~\cite{Catani:1996vz, Catani:2002hc, Dittmaier:1999mb, Dittmaier:2008md, Kallweit:2017khh}, as implemented in \OpenLoops (see 
Ref.~\cite{Buccioni:2019sur}, Section~3.4). At two-loop, infrared divergences are instead regulated via a fictitious photon mass. Differences between dimensional regularisation and mass regularisation are found to be below the percent level at one-loop in LA.  

We consider the following one- and two-loop NLL results in LA, 
\begin{align}
&\NLLpone = (1+ \hat\delta^{(1)}_{\mathrm{LL}} + \hat\delta^{(1)}_{\mathrm{NLL}}  +  \hat\delta^{(1)}_{\mathrm{NNLL}} + \mathbf{I}_{1}^{\rm QED})\, \LO \,,\\
&\text{NNLO LL+NLL EW}= (1+ \hat\delta^{(2)}_{\mathrm{LL}} + \hat\delta^{(2)}_{\mathrm{NLL}})\, \LO\,,
\end{align}
i.e.\ at one-loop we include angular-dependent NNLL contributions. These contributions cannot be reliably controlled in LA. However, they can become numerically relevant for $r_{jk} \gg r_{j'k'}$, i.e.\ in phase-space regions where not all kinematic invariants are of the order of the centre-of-mass energy $s$. 

Besides these purely logarithmic corrections, we also consider the combination of exact virtual NLO EW corrections with virtual NNLO corrections in LA, 
\begin{align}
 \nNLO  = \NVI + (\hat\delta^{(2)}_{\mathrm{LL}} + \hat\delta^{(2)}_{\mathrm{NLL}})\, \LO.
\end{align}
This combination can be regarded as the best available prediction for higher-order EW effects, which is now readily available for a large range of processes.

Results for the LHC at $\sqrt{S}=13~\mathrm{TeV}$ are obtained using \Sherpa~\cite{Sherpa:2019gpd} 
for phase-space integration and \OpenLoops as the amplitude provider. We use the input parameters
\begin{align}
 m_Z&=91.1876~\mathrm{GeV}\,, \quad m_W=80.399~\mathrm{GeV}\,, \nonumber\\
 m_H&=125~\mathrm{GeV}\,, \qquad \quad m_{\mathrm{t}}=173.2~\mathrm{GeV}\,, \nonumber\\
 \Gamma_Z&= \Gamma_W=0~\mathrm{GeV}\,, \quad G_\mu =1.16637 \times 10^{-5}~\mathrm{GeV}^{-2}\,,
\end{align}
with all other quarks and leptons treated as massless. The EW regularisation scale is set 
to $\muew = \sqrt{\hat{s}}$. We use the NNPDF3.0 PDF 
set~\cite{NNPDF:2014otw} via LHAPDF6~\cite{Buckley:2014ana}, with $\mu_R=\mu_F=H_T/2$, 
where $H_T$ is the scalar sum of final-state transverse momenta; all relative EW 
corrections are independent of this scale choice. Jets are clustered with the anti-$k_T$ 
algorithm~\cite{Cacciari:2008gp} as implemented in \texttt{FastJet}~\cite{Cacciari:2011ma}, 
with $R=0.4$, $p_{{\rm T},j}> 30~\mathrm{GeV}$ and $| y_{j} | < 4.4$. For final-state vector bosons we require $| \eta_{V} | < 3$ in order to mimic experimental cuts on leptonic decay products. 
All processes are otherwise considered fully inclusive, and partonic events are analysed 
with \Rivet~\cite{Bierlich:2019rhm}.

We present one- and two-loop EW results for a representative selection of LHC process classes 
involving massless fermions and transverse gauge bosons: $V+$jets, \linebreak $VV+$jets, and $VVV$, 
chosen to highlight different features and limitations of the LA.

%%%%%%%%%%%%%%%%%%%%%%%%%%%%%%%%%%%%%%%%%%%%%%%
\subsubsection*{$\mathbf{W}$+jet}
%%%%%%%%%%%%%%%%%%%%%%%%%%%%%%%%%%%%%%%%%%%%%%%

\begin{figure*}[tbp]
\centering
	\includegraphics[width=\setrelwidth\textwidth]{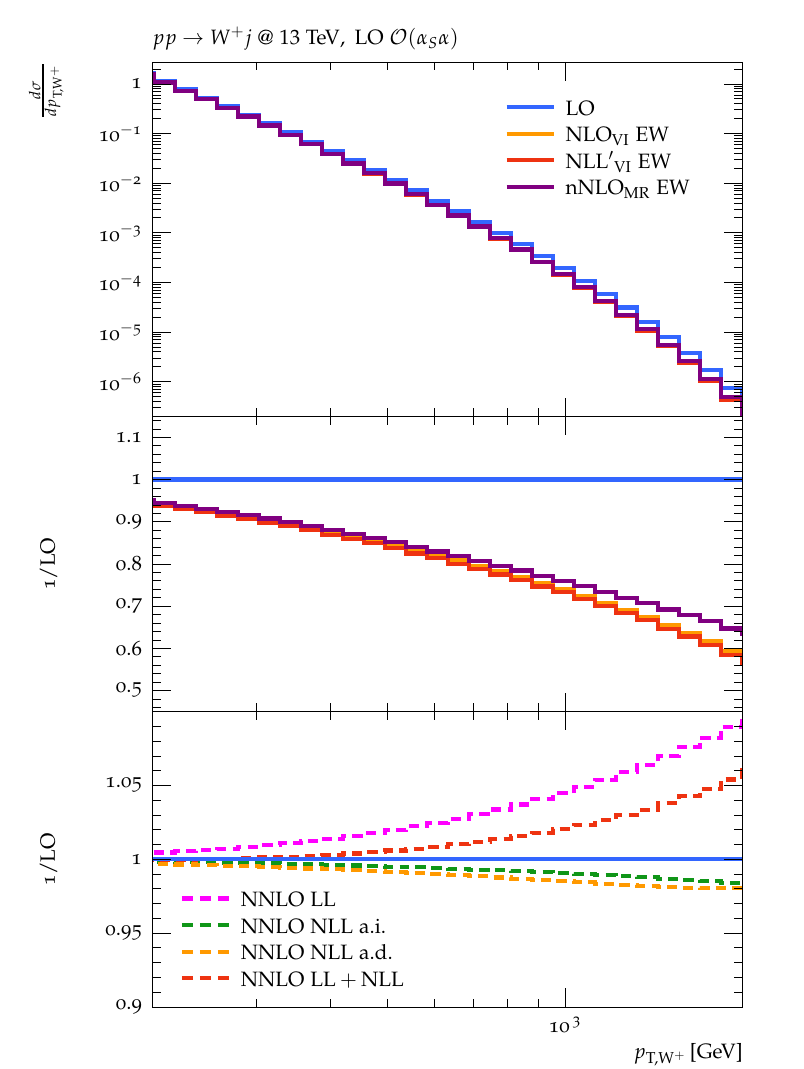}
	\includegraphics[width=\setrelwidth\textwidth]{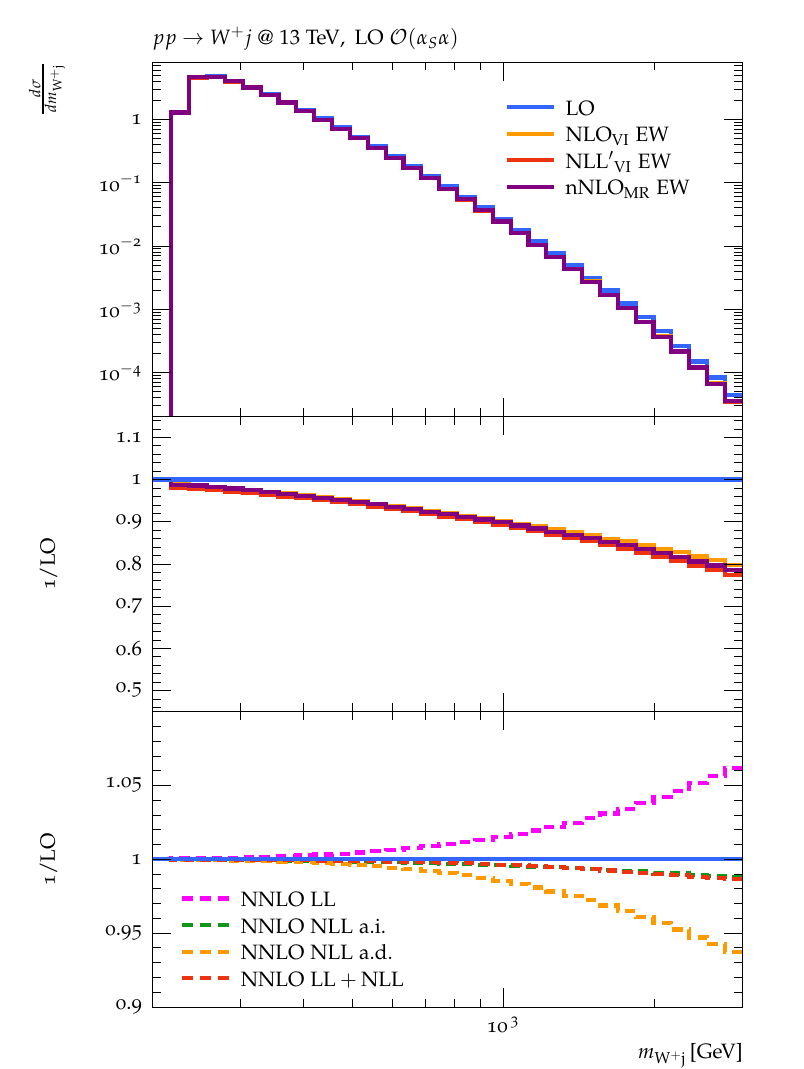}
\caption{Differential distributions in the transverse momentum of the 
$W^+$~boson $p_{\rm{T},W^+}$ (left) and in the invariant mass of the $W^+j$ system $m_{\rm{W^+j}}$  (right) 
 in $pp \to W^+j$ at $\sqrt{s}=13~\rm{TeV}$. The first panel shows absolute predictions at LO~(blue), \NVI~(orange), where the subscript ``VI'' indicates that the QED $I$-operator is added to the virtual amplitudes in DR, \NLLp~(red), and \nNLO~(purple). The second panel shows the corresponding relative corrections with respect to LO. The third panel shows the various contributions at \NLLtwo in MR, as discussed in Section~\ref{sec:two_loop_factorisable} and \ref{sec:two_loop_ai_nll}, normalised to LO:
 \LL (dashed magenta),  \NLLai (dashed green),  \NLLad (dashed orange) and full LL+NLL (dashed red). 
}
\label{fig:wj}
\end{figure*}

In Fig.~\ref{fig:wj} we show the transverse-momentum $p_{\mathrm{T},W^+}$ and $W^+j$ invariant-mass distributions for $W^{+}+\text{jet}$ production. 
The $p_{\mathrm{T,W^+}}$ distribution (left panel in Fig.~\ref{fig:wj}) exhibits logarithmically enhanced corrections at both one- and two-loop level. These corrections yield up to $-35\%$ with respect to LO in the TeV range. The two-loop corrections are positive and amount to up to $+5$--$6\%$ at $p_{\mathrm{T,W^+}}=2~\text{TeV}$, thus reducing the size of the corrections observed at one-loop. 
This is consistent with the two-loop EW predictions for $W^{+}+\text{jet}$ production reported in Refs.~\cite{Kuhn:2007cv,Lindert:2017olm}.

Comparing the individual contributions at two-loop, we observe a clear hierarchy: NLL corrections (both \linebreak angular-dependent and angular-independent) are negative and about a factor of five smaller than the corresponding positive \LL contribution. As a result, the total two-loop correction (shown as the dashed red curve in the lower ratio plot) is largely dominated by the LL corrections.

The right panel of Fig.~\ref{fig:wj} shows the invariant-mass distribution of the $W^+j$ system. In this case, the LA condition of Eq.~\eqref{la} is not fully satisfied. As a result, the logarithmic hierarchy is spoilt, and at two-loop the angular-dependent NLL corrections largely compensate the \LL contributions, leading to a total two-loop EW correction of only about $-2\%$ at $m_{\rm{W^+j}}=3~\text{TeV}$, whereas the \LL corrections alone reach $+6\%$.

%%%%%%%%%%%%%%%%%%%%%%%%%%%%%%%%%%%%%%%%%%%%%%%
\subsubsection*{$\mathbf{W+2}$ jets}
\begin{figure*}[tbp]
\centering
	\includegraphics[width=\setrelwidth\textwidth]{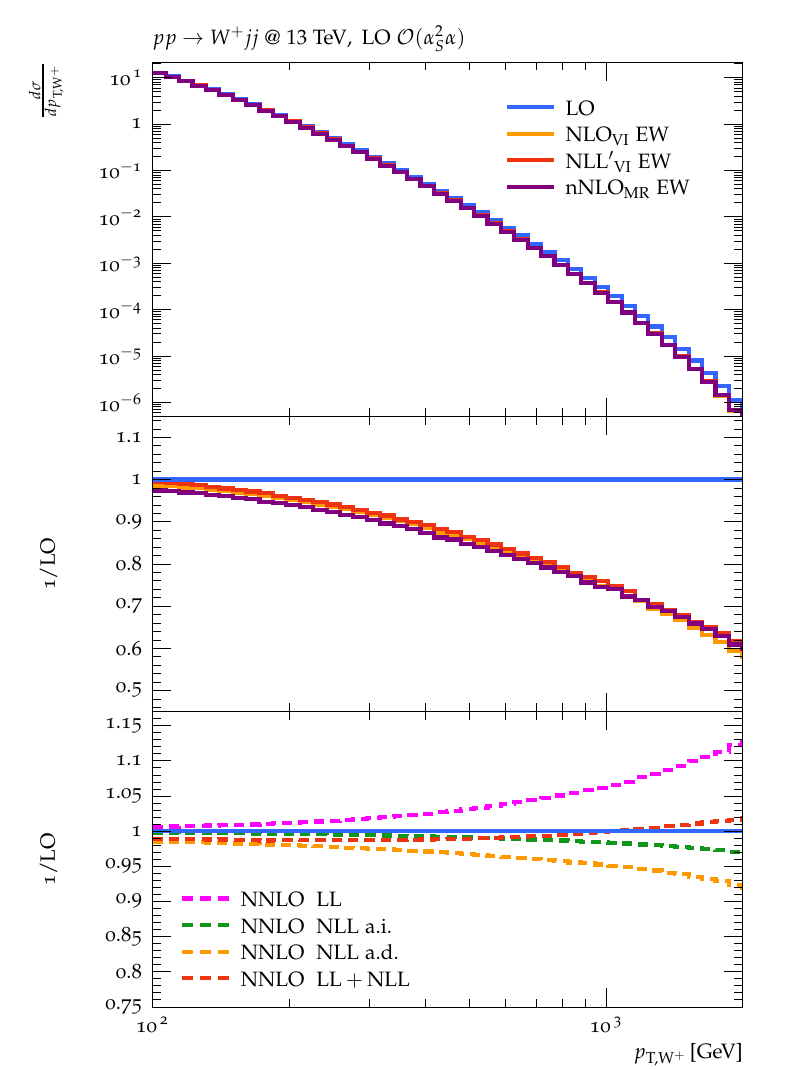}
	\includegraphics[width=\setrelwidth\textwidth]{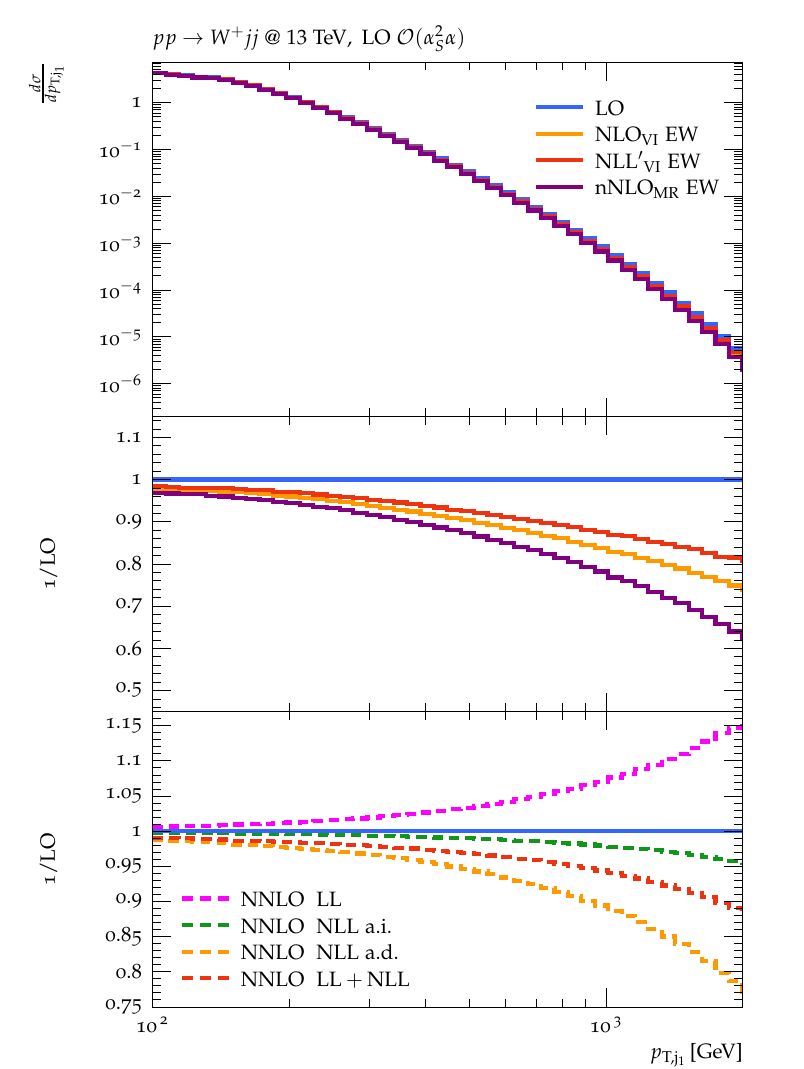}
\caption{Differential distributions in the transverse momentum of the 
$W^+$~boson $p_{\rm{T,W^+}}$ (left) and of the hardest jet $p_{\rm{T,j_1}}$  (right) 
 in $pp \to W^+jj$ at $\sqrt{s}=13~\rm{TeV}$. Curves as in Fig.~\ref{fig:wj}.}
\label{fig:wjj_pT}
\end{figure*}

%%%%%%%%%%%%%%%%%%%%%%%%%%%%%%%%%%%%%%%%%%%%%%%
Next, we consider EW corrections to $W^{+} + 2$ jet production at $\ord(\as^{2}\alpha)$. Corresponding differential distributions in $p_{\mathrm{T,W^+}}$ and $p_{\mathrm{T,j_1}}$ are shown in Fig.~\ref{fig:wjj_pT}.

For $p_{\mathrm{T,W^+}}$ (left), the relative EW corrections exhibit a pattern and quantitative impact similar to those observed in $W^{+}+\text{jet}$ production.  In the $W^{+} + 2$ jets case, however, the angular-dependent NLL corrections are twice as large as the angular-independent ones, \NLLai, reaching $-8\%$ at $p_{\mathrm{T,W^+}}=2~\text{TeV}$. As a consequence, the overall two-loop LL+NLL correction is reduced, amounting to about $2\%$ at the same value of the $W^+$ transverse momentum.

Considering $p_{\mathrm{T,j_1}}$ (right) in the inclusive phase-space selection considered here, we observe a violation of the LA condition~\eqref{la}. In fact, at large $p_{\mathrm{T,j_1}}$ the angular-dependent NLL contributions exceed the \LL term, driving the total two-loop correction to $-11\%$ at $p_{\mathrm{T,j_1}}=2~\text{TeV}$. The total two-loop corrections include the angular-independent NLL contributions, which amount to approximately $-4\%$ at the same transverse momentum. As noted in Ref.~\cite{Lindert:2023fcu} in the context of $Z + 2$ jet production, such a violation of the LA condition originates from the fact that at
large $p_{\mathrm{T,j_1}}$ the recoil of the hardest jet can be absorbed by
the second jet, while the vector boson can remain soft. Such configurations yield a separation of scales in violation of Eq.~\eqref{la}.

\begin{figure*}[tbp]
\centering
	\includegraphics[width=\setrelwidth\textwidth]{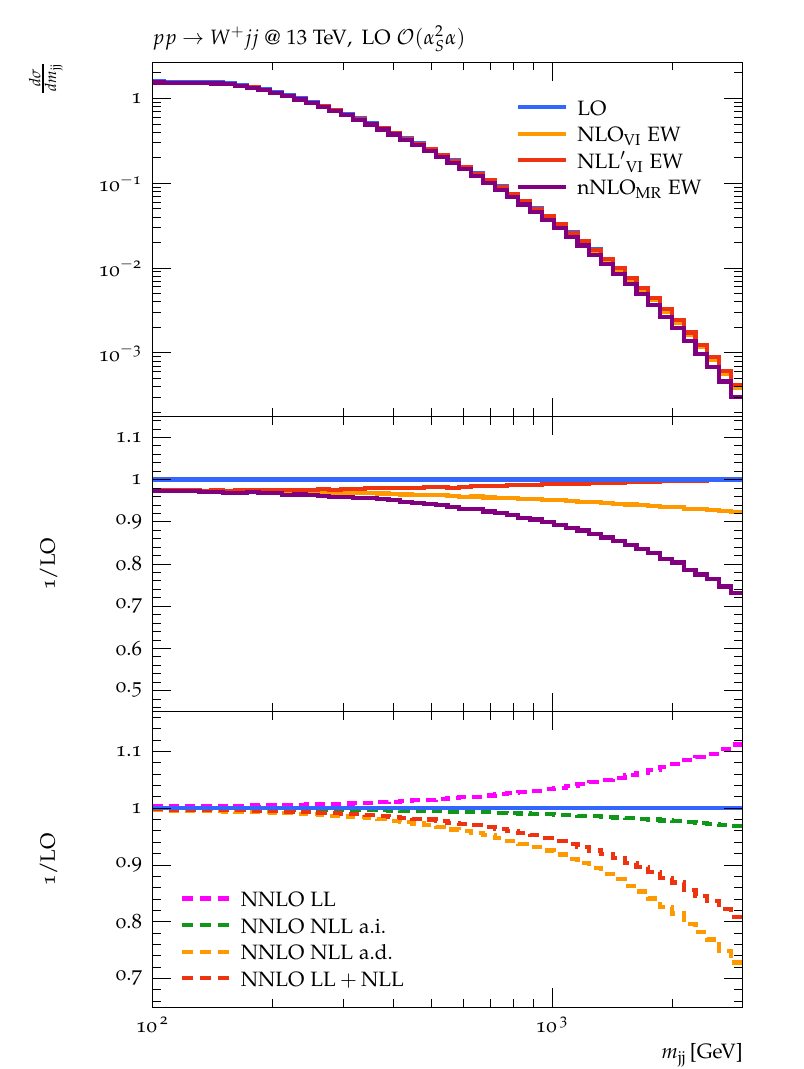}
	%plots/hadron_level_comb/wjj/m_jj_log.pdf}
	\includegraphics[width=\setrelwidth\textwidth]{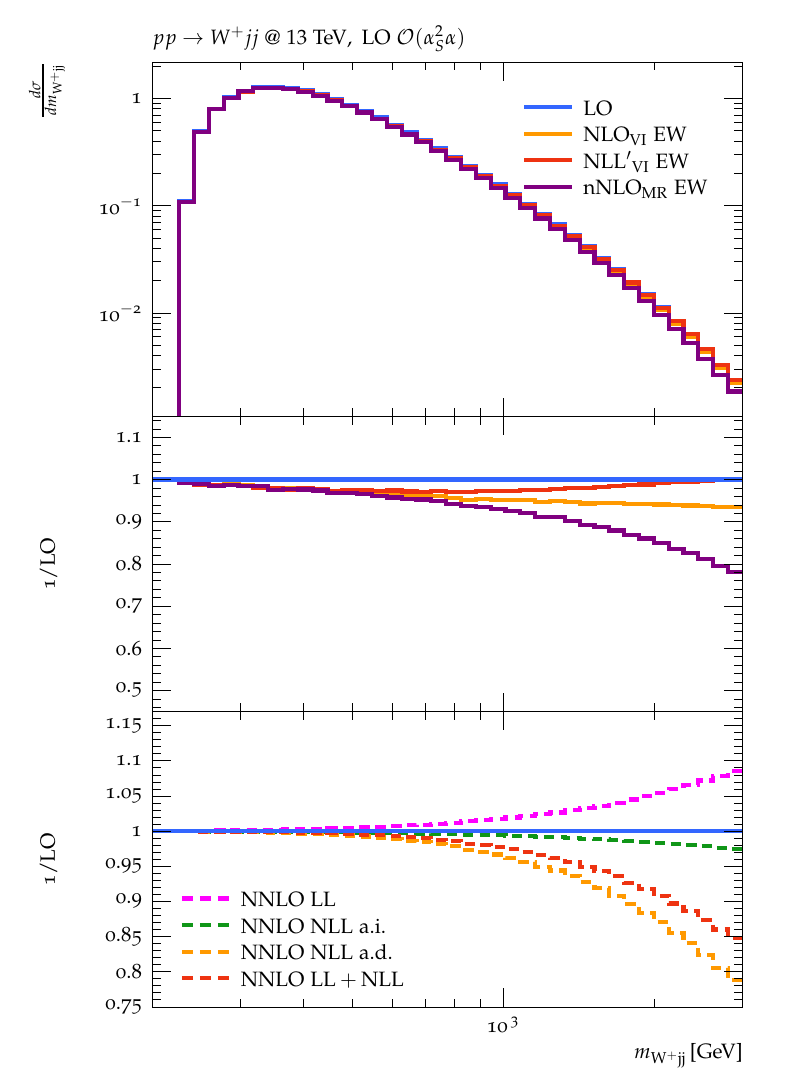}
	\caption{Differential distributions in the invariant mass $m_{\rm{jj}}$ (left) and $m_{\rm{W^+jj}}$ (right)
 in $pp \to W^+jj$ at $\sqrt{s}=13~\rm{TeV}$. Curves as in Fig.~\ref{fig:wj}.}
\label{fig:wjj_mass}
\end{figure*}

In Fig.~\ref{fig:wjj_mass}, we present the invariant-mass distributions of the di-jet system (left) and of the full $W^+jj$ system (right). In both cases, the LA condition~\eqref{la} is violated. This originates from the fact that large invariant masses can be realised through forward configurations, which are subject to a significant separation of scales. As a result, at two-loop level, sizeable positive \LL corrections of around $10\%$ at $m_{\rm{jj}}=3$~TeV and $m_{\rm{W^+jj}}=3$~TeV are overcompensated by larger negative angular-dependent \NLLad contributions, amounting to $-28\%$ and $-22\%$, respectively. Combined with the angular-independent NLL contributions of about $-3\%$ in both distributions, the total two-loop LL+NLL correction amounts to $-20\%$ at $m_{\rm{jj}}=3$~TeV and $-15\%$ at $m_{\rm{W^+jj}}=3$~TeV.

%%%%%%%%%%%%%%%%%%%%%%%%%%%%%%%%%%%%%%%%%%%%%%%
\subsubsection*{$\mathbf{ZZ}$}
%%%%%%%%%%%%%%%%%%%%%%%%%%%%%%%%%%%%%%%%%%%%%%%

\begin{figure*}[tbp!]
\centering
	\includegraphics[width=\setrelwidth\textwidth]{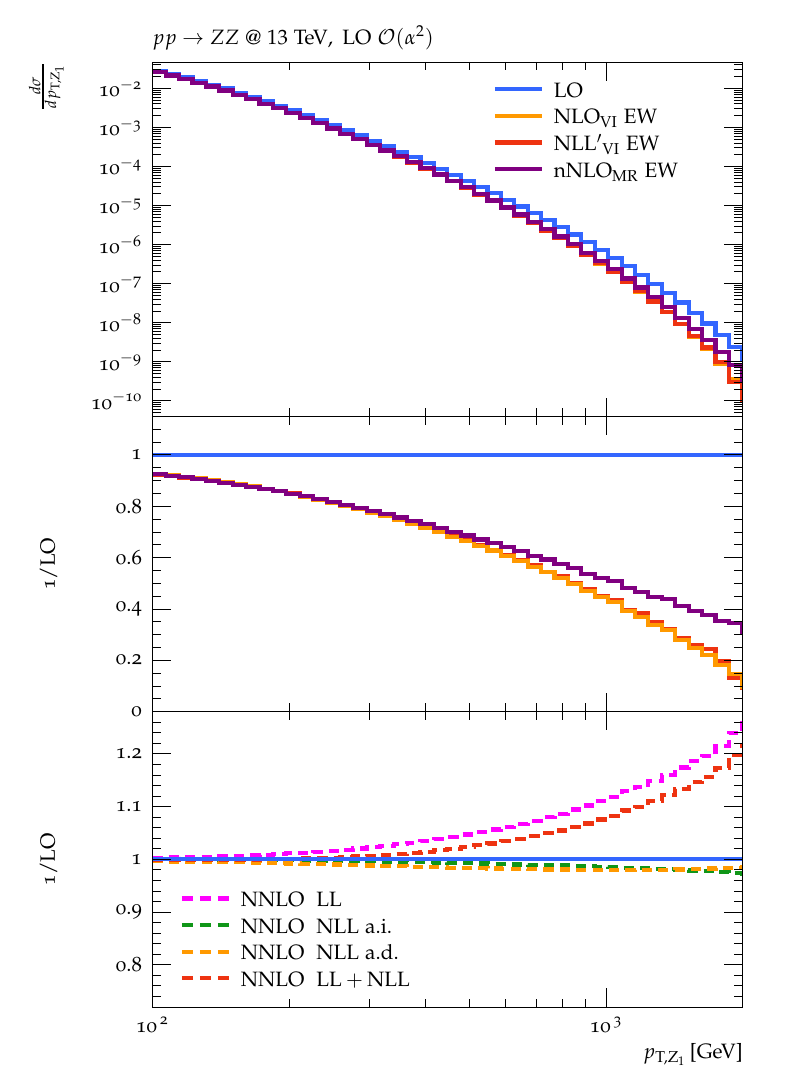}
	\includegraphics[width=\setrelwidth\textwidth]{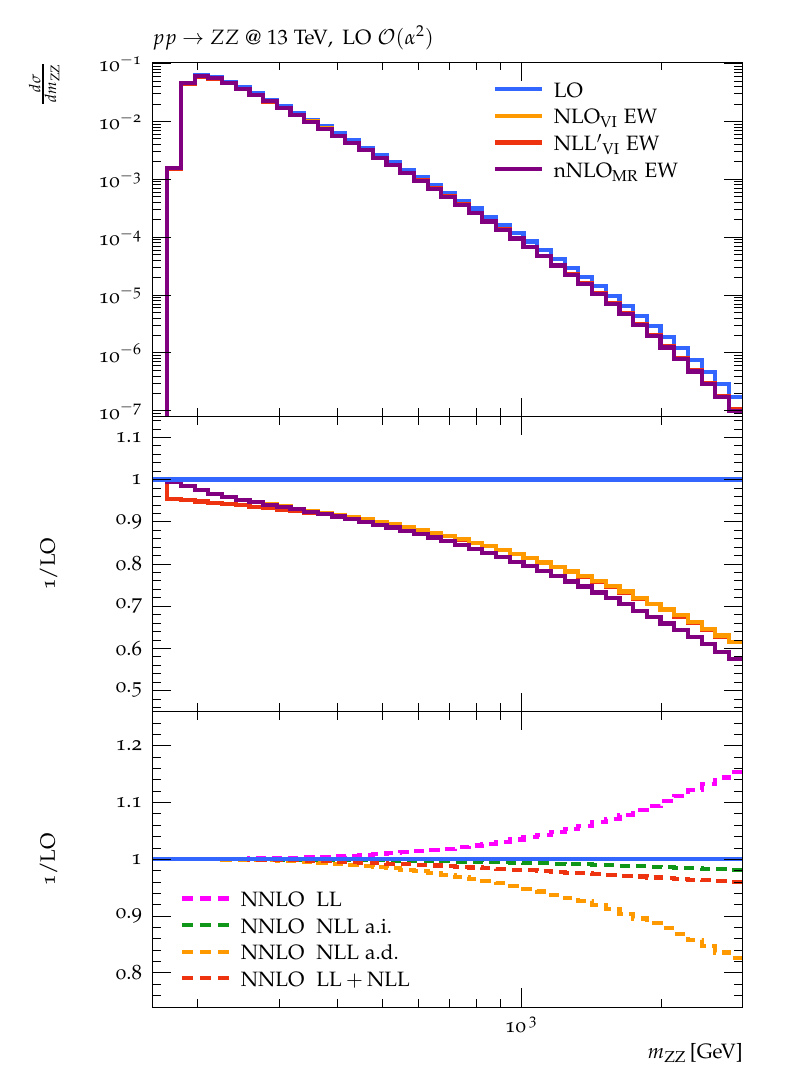}
\caption{Differential distributions in the transverse momentum of the 
$Z$~boson $p_{\rm{T,Z}}$ (left) and in the invariant mass of the $ZZ$ system $m_{\rm{ZZ}}$  (right) 
 in $pp \to ZZ$ at $\sqrt{s}=13~\rm{TeV}$. Curves as in Fig.~\ref{fig:wj}.}
\label{fig:zz}
\end{figure*}

In Fig.~\ref{fig:zz} we consider hadronic $ZZ$ production and present the transverse-momentum distribution of one of the back-to-back  $Z$ bosons (left panel) and the invariant-mass distribution of the $ZZ$ system (right panel). In the transverse-momentum distribution we observe a hierarchical Sudakov-like behaviour: the \LL contribution drives the total two-loop correction. The total two-loop LL+NLL correction reaches $+22\%$ at $p_{\rm{T,Z}}=2$~TeV.
Both angular-independent and angular-dependent NLL corrections remain below $-2\%$ each.

In contrast, in the invariant-mass distribution of the $ZZ$ system (right panel), we observe large angular- \linebreak dependent corrections despite the pseudorapidity constraint imposed on the $Z$~bosons. The angular-dependent NLL corrections overcompensate the LL corrections, yielding 
a total two-loop correction of up to only $-4\%$ in the tail of this distribution.

%%%%%%%%%%%%%%%%%%%%%%%%%%%%%%%%%%%%%%%%%%%%%%%
\subsubsection*{$\mathbf{ZZ+}$jet}
%%%%%%%%%%%%%%%%%%%%%%%%%%%%%%%%%%%%%%%%%%%%%%%

\begin{figure*}[tbp]
\centering
	\includegraphics[width=\setrelwidth\textwidth]{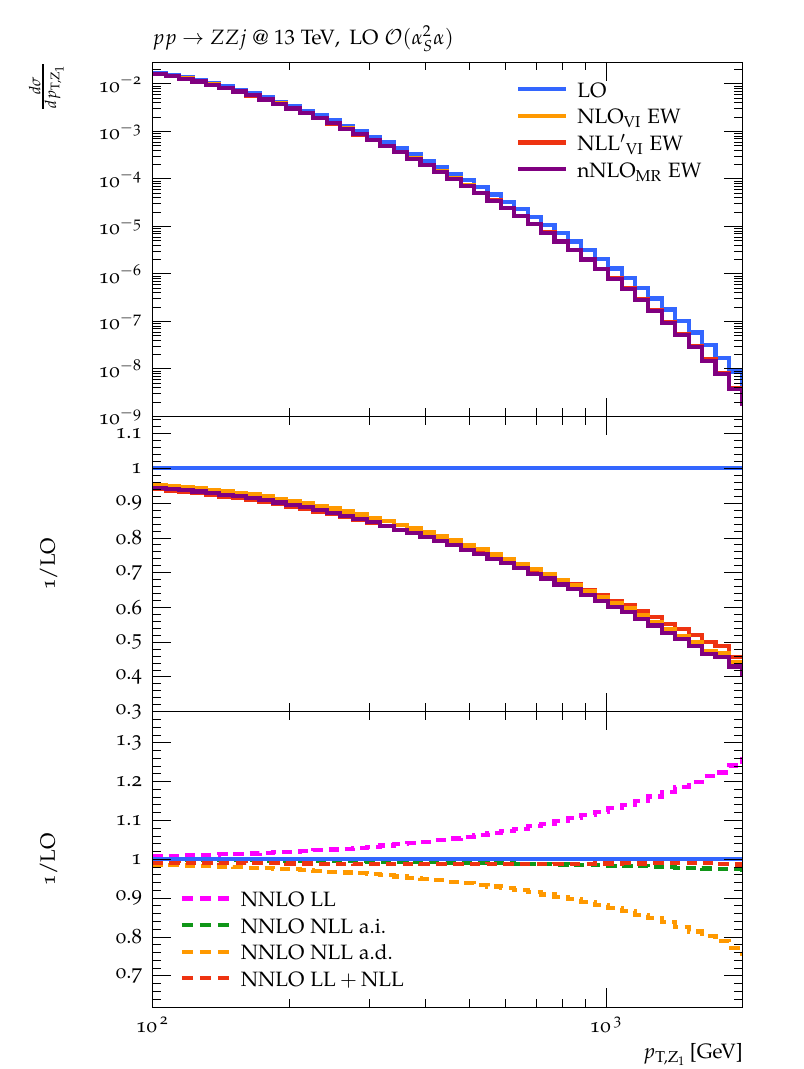}
	\includegraphics[width=\setrelwidth\textwidth]{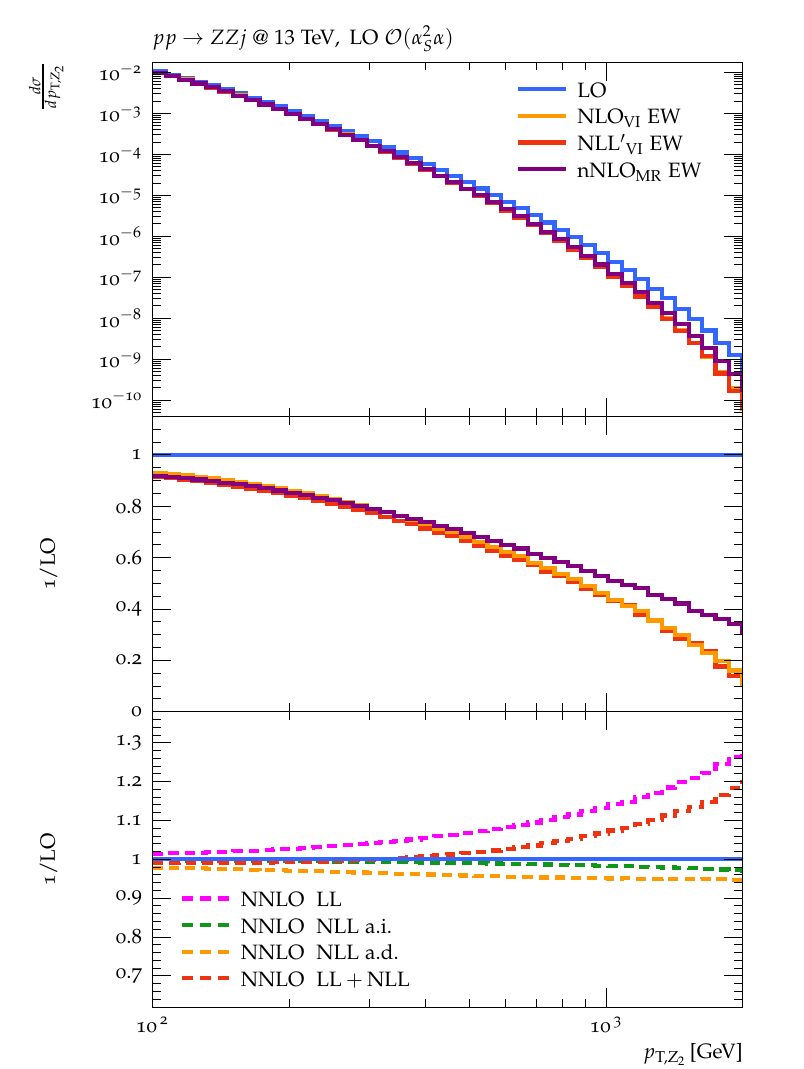}
\caption{Differential distributions in the transverse momentum of the harder, $p_{\rm{T},Z_1}$ (left), and softer, $p_{\rm{T},Z_2}$ (right), $Z$ bosons 
 in $pp \to ZZj$ at $\sqrt{s}=13~\rm{TeV}$. Curves as in Fig.~\ref{fig:wj}.}
\label{fig:zzj_z}
\end{figure*}

Next, we consider $ZZ+$jet production. In Fig.~\ref{fig:zzj_z} we display the transverse-momentum distribution of the harder $Z$ boson $p_{\rm{T,Z_1}}$ (left) and the softer $Z$ boson $p_{\rm{T,Z_2}}$ (right). In the $p_{\rm{T,Z_1}}$ distribution we observe large \LL contributions (up to +25\% at $p_{\rm{T,Z_1}}=2$~TeV) which are almost exactly compensated by equally large angular-dependent NLL corrections, resulting in total two-loop EW corrections of only a few percent in the multi-TeV regime. This violation of the hierarchical structure of the logarithmic EW corrections originates from contributions where the hardest $Z$ boson recoils against a jet, while the second $Z$ boson remains soft. The very same dynamics generate giant K-factors at NLO QCD, see e.g. Ref. \cite{Grazzini:2019jkl}.

In contrast, in the transverse-momentum distribution of the softer $Z$ boson $p_{\rm{T,Z_2}}$ we observe a hierarchical structure, where the LL dominates the total two-loop corrections and NLL effects remain at the few percent level. At $p_{\rm{T,Z_2}}=2$~TeV the total two-loop corrections reach $+20\%$, very similar to the $p_{\rm{T,Z_1}}=p_{\rm{T,Z_2}}$ distribution in $ZZ$ production.

\begin{figure*}[tbp]
\centering
	\includegraphics[width=\setrelwidth\textwidth]{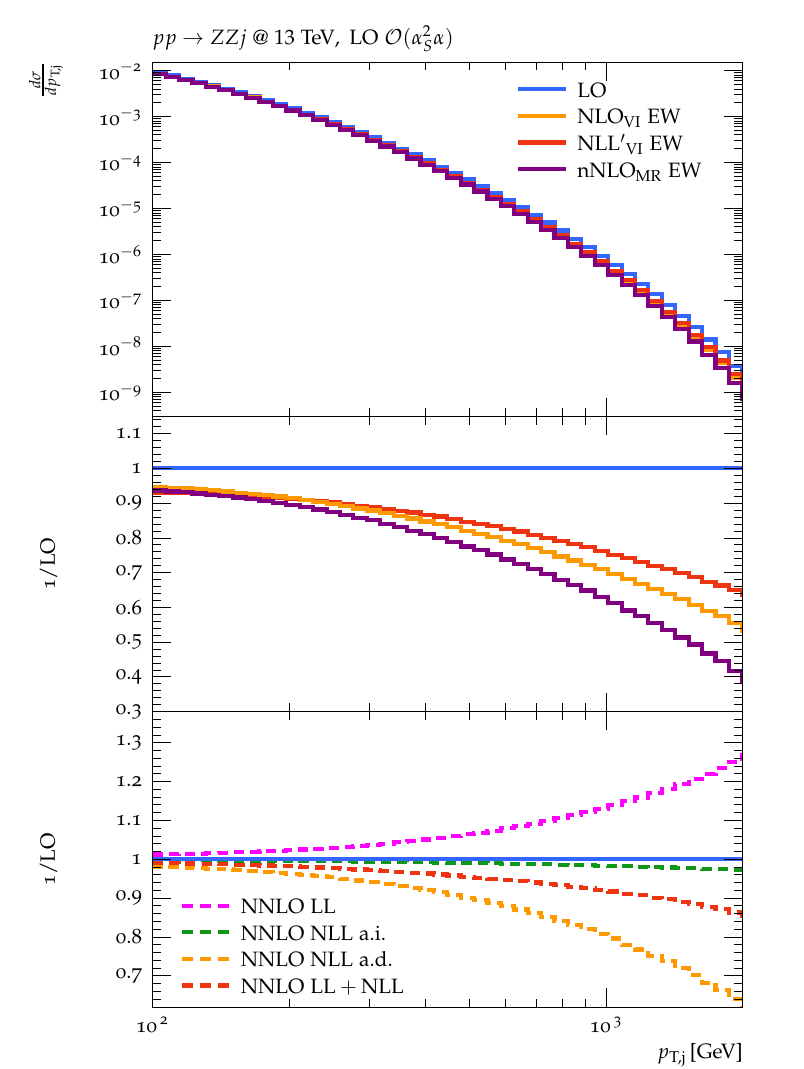}
		\includegraphics[width=\setrelwidth\textwidth]{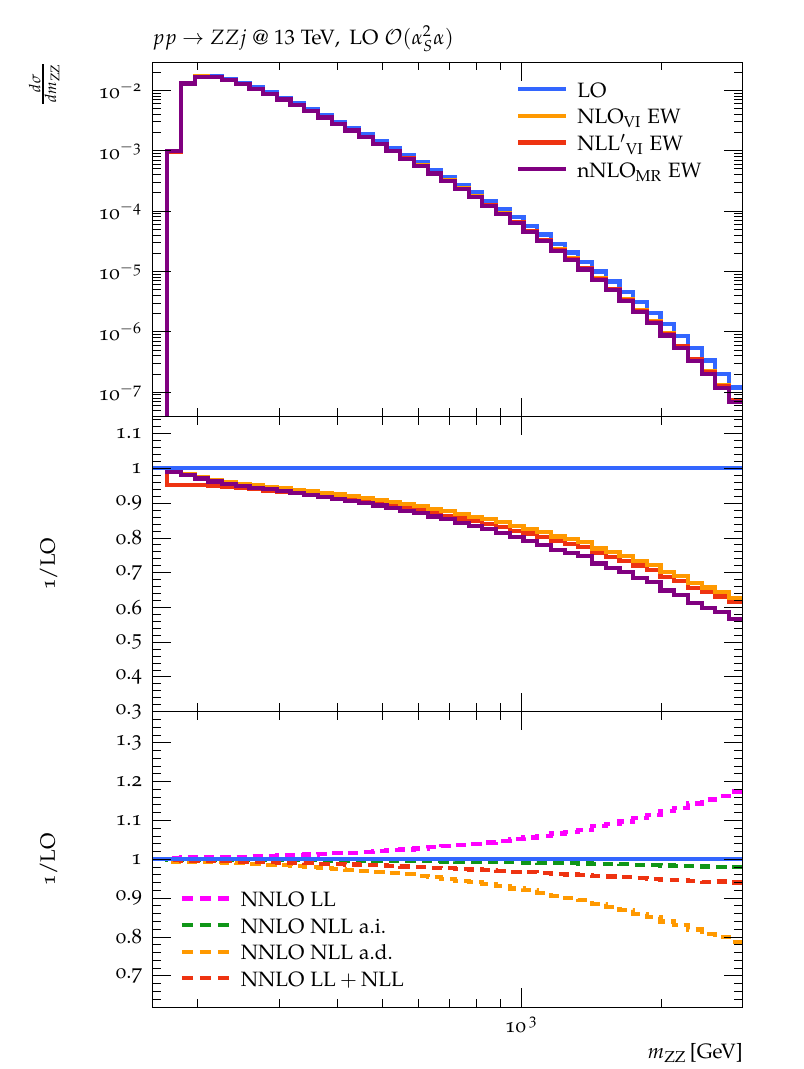}
\caption{Differential distributions in the transverse momentum of the jet $p_{\rm{T,j}}$  (left) and in the invariant mass of the  $ZZ$ system $m_{\rm{ZZ}}$ (right)
 in $pp \to ZZj$ at $\sqrt{s}=13~\rm{TeV}$. Curves as in Fig.~\ref{fig:wj}.}
\label{fig:zzj_j}
\end{figure*}

In Fig.~\ref{fig:zzj_j} we display the transverse-momentum distribution of the jet $p_{\rm{T,j}}$ (left) and the  invariant-mass distribution of the $ZZ$ system (right).
 In the $p_{\rm{T,j}}$ distribution we observe, as already discussed in Ref.~\cite{Lindert:2023fcu} in the context of one-loop EW corrections in LA, that the assumptions of the LA are violated. As a result, at two-loop level a large negative \NLLad correction of about $-36\%$ at $p_{\rm{T,j}}=2$~TeV overcompensates the \LL contribution, which yields a $28\%$ correction at the same transverse momentum. Including \NLLai contributions, the overall two-loop correction reaches up to $-14\%$ in the tail of this distribution. We observe a similar behaviour in the invariant-mass distribution of the $Z_1j$ system. Here, the total two-loop correction at LL+NLL accuracy reaches $-15\%$ in the tail.

Considering the invariant-mass distribution of the $ZZ$ system (right plot in Fig.~\ref{fig:zzj_j}), we observe that the positive \LL correction is overcompensated by the large negative angular-dependent NLL contribution. Angular-independent \NLL corrections remain small, amounting to just $-2\%$ in the tail of the distribution. The total two-loop EW correction at LL+NLL accuracy reaches $-6\%$ at $m_{\rm{ZZ}}=3$~TeV.

%%%%%%%%%%%%%%%%%%%%%%%%%%%%%%%%%%%%%%%%%%%%%%%
\subsubsection*{$\mathbf{ZZ\gamma}$}
%%%%%%%%%%%%%%%%%%%%%%%%%%%%%%%%%%%%%%%%%%%%%%%

\begin{figure*}[tbp]
\centering
	\includegraphics[width=\setrelwidth\textwidth]{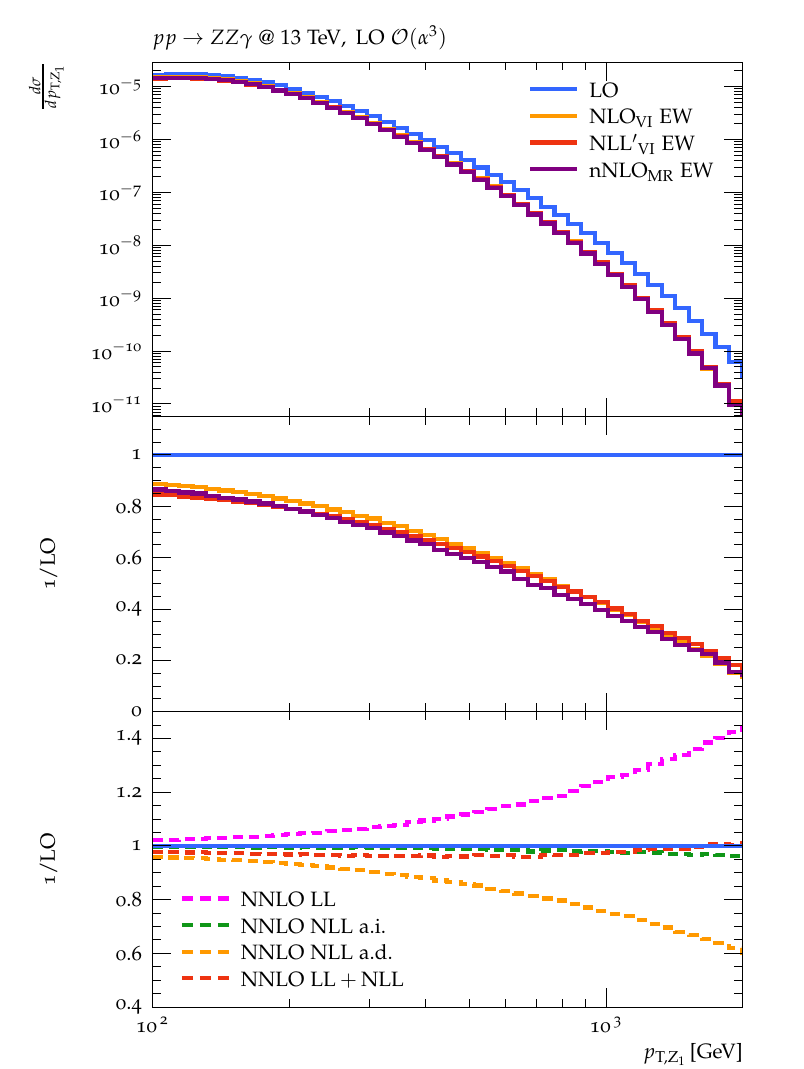}
	\includegraphics[width=\setrelwidth\textwidth]{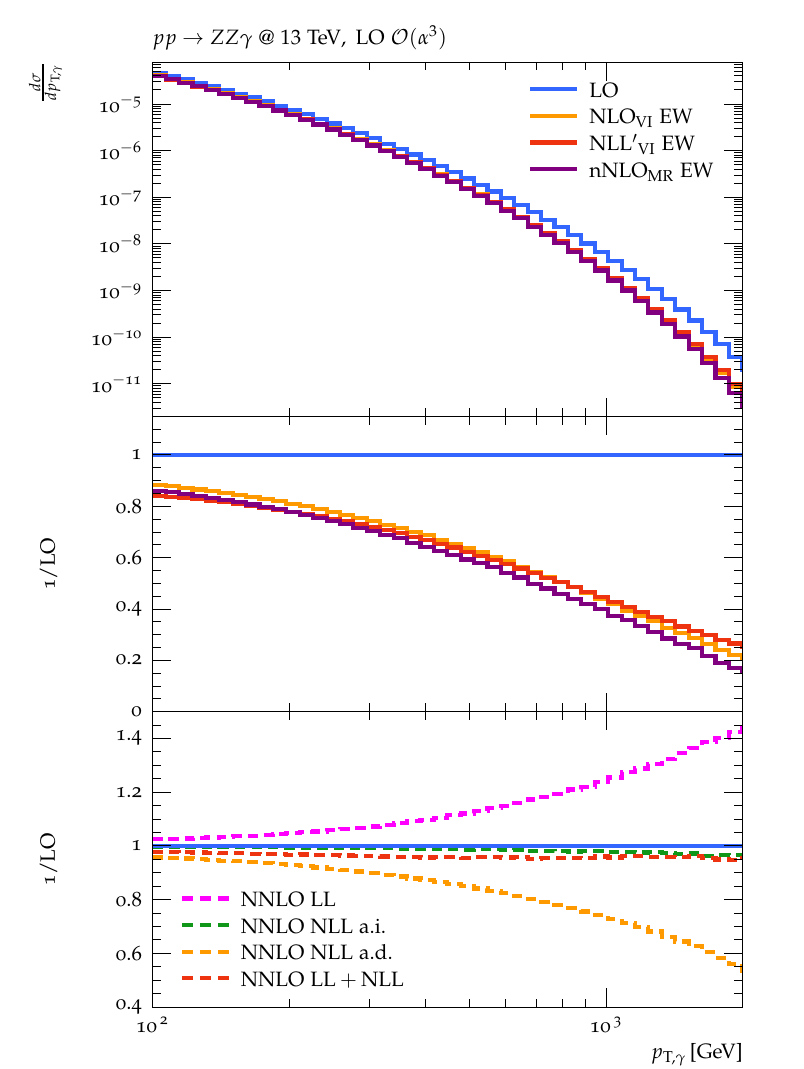}
\caption{Differential distributions in the transverse momentum of the hardest
$Z$ boson $p_{\rm{T},Z_1}$ (left) and of the photon $p_{\rm{T},\gamma}$ (right)
 in $pp \to ZZ\gamma$ at $\sqrt{s}=13~\rm{TeV}$. Curves as in Fig.~\ref{fig:wj}.}
\label{fig:zza_pt}
\end{figure*}

\begin{figure*}[tbp]
\centering
	\includegraphics[width=\setrelwidth\textwidth]{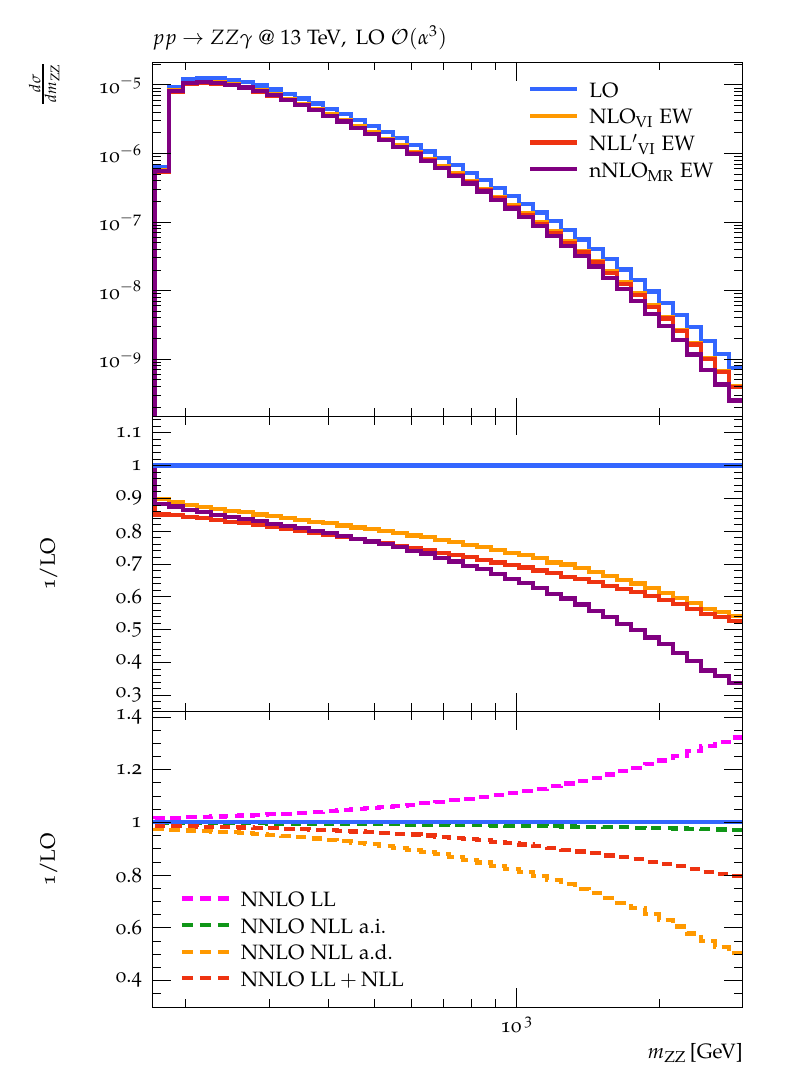}
	\includegraphics[width=\setrelwidth\textwidth]{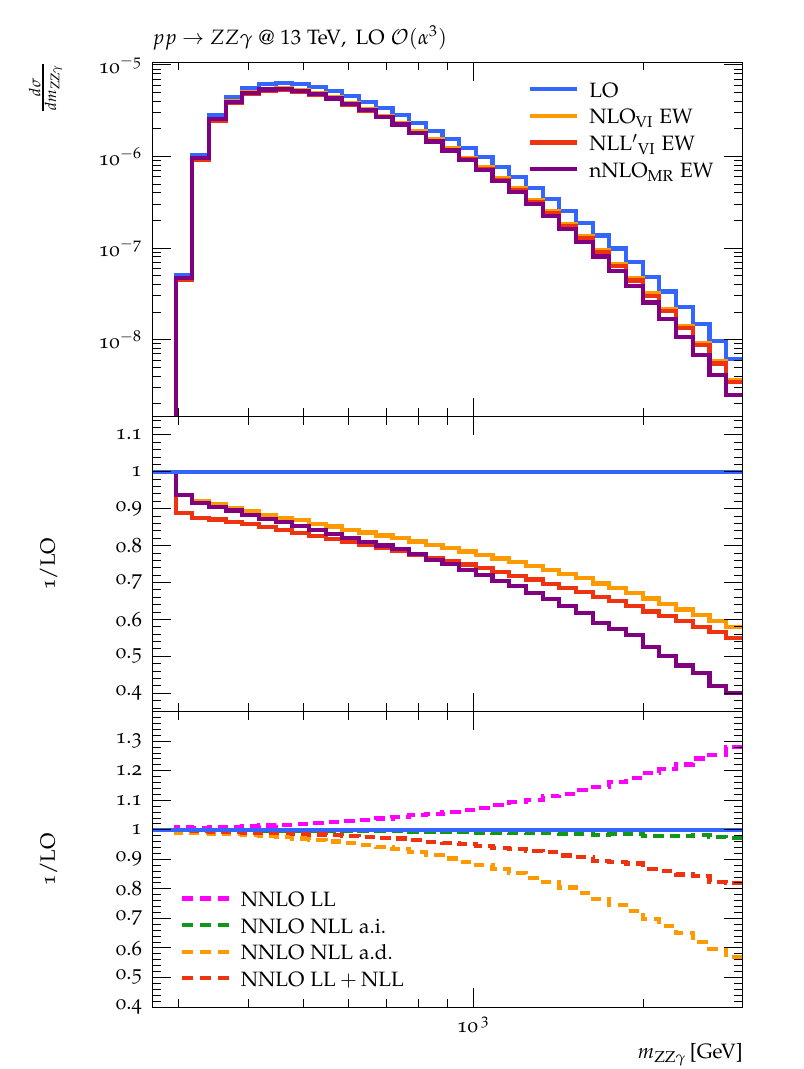}
\caption{Differential distributions in the invariant mass of the $ZZ$ system  $m_{\rm{ZZ}}$   (left) and of the $ZZ\gamma$ system $m_{\rm{ZZ\gamma}}$ (right)
 in $pp \to ZZ\gamma$ at $\sqrt{s}=13~\rm{TeV}$. Curves as in Fig.~\ref{fig:wj}.}
\label{fig:zza_m}
\end{figure*}

Finally, we turn to the $pp \to ZZ\gamma$ process, which has only recently been observed at the LHC~\cite{ATLAS:2026zgd}. We consider this process subject to the requirement $p_{\rm T, \gamma} > 100\,$GeV on the photon. This cut avoids  a Born-level infrared singularity.
In Fig.~\ref{fig:zza_pt} we show the transverse-momentum distribution of the hardest $Z$~boson (left) and of the photon (right). In both distributions the LL corrections amounts to up to $+35\%$, which gets largely compensated by equally large angular dependent NLL corrections resulting in complete two-loop corrections of only a few percent in the multi-TeV regime. 
We observe a similar large compensation between LL and NLL angular dependent corrections in the invariant mass of the $ZZ$ pair and of the entire $ZZ\gamma$ system shown in Fig.~\ref{fig:zza_m}. In fact, the NLL angular dependent corrections overcompensate the LL effects resulting in overall two-loop corrections of around $-20\%$ in the tails of both kinematic distributions.

%%%%%%%%%%%%%%%%%%%%%%%%%%%%%%%%%%%%%%%%%%%%%%%
\section{Conclusions} \label{sec:conclusions}
%%%%%%%%%%%%%%%%%%%%%%%%%%%%%%%%%%%%%%%%%%%%%%%
We have presented the implementation of two-loop electroweak corrections in the
logarithmic approximation at NLL accuracy in the amplitude generator \OpenLoops. The
implementation covers processes involving an arbitrary number of massless fermions and
transversely polarised gauge bosons, and is fully process-independent and largely
model-independent. It is based on the diagrammatic approach of
Refs.~\cite{Denner:2003wi,Denner:2006jr} for the angular-dependent two-loop
logarithms, combined with the expanded all-order resummation results of
Refs.~\cite{Melles:2000gw,Melles:2001mr} for the angular-independent contributions.
The implementation extends the one-loop EW Sudakov framework of
Ref.~\cite{Lindert:2023fcu} to two loops via the same pseudo-counterterm approach,
which provides a \linebreak streamlined bookkeeping of the required $\text{SU}(2)$-correlated
Born matrix elements with up to four soft vector boson insertions. A key feature of
the construction is based on the observation in~Refs. \cite{Denner:2003wi,Denner:2006jr}
that the two-loop LL and angular-dependent NLL contributions reduce, after
Ward-identity cancellations, to products of one-loop $D_0$ functions and
$\beta$-function coefficients, making the two-loop implementation both compact and
efficient.

We have validated the implementation through detailed comparisons of the individual
LL, NLL a.i.\ and NLL a.d.\ logarithmic coefficients against analytical results
available in the literature~\cite{Denner:2006jr,Kuhn:2004em,Kuhn:2005gv,Kuhn:2007qc} for a
representative set of partonic processes. 
Excellent agreement has been found for all contributions and all non-mass-suppressed
helicity configurations. 

We have presented numerical results for a representative set of LHC processes,
including $V+$jets, $VV+$jets, and $VVV$ production, illustrating both the
quantitative impact and the limitations of the EW two-loop corrections in LA. 
For observables where all kinematic invariants are of comparable magnitude --
such as transverse-momentum distributions of the vector bosons -- the two-loop
corrections exhibit a clear logarithmic hierarchy, with the LL contributions dominating
and NLL corrections remaining sub-dominant. In these cases, the two-loop corrections
are positive and amount to several percent in the TeV range, partially compensating the
large negative one-loop Sudakov corrections. However, for invariant-mass distributions
and observables sensitive to forward or unbalanced configurations, the LA condition can
be violated due to the appearance of widely separated kinematic scales. In such
regions, angular-dependent NLL contributions can become comparable to or even exceed
the LL terms, spoiling the expected logarithmic hierarchy and, in some cases,
overcompensating the LL corrections. 

The implementation presented here provides a practical and automated tool for the
inclusion of two-loop EW Sudakov corrections in precision predictions for LHC
processes. At high energies, where the  overall EW corrections can reach several tens of
percent, the two-loop contributions computed here are of order a few percent and
therefore non-negligible for precision phenomenology. Their systematic inclusion is
essential to reduce theoretical uncertainties from missing higher-order EW corrections
in the tails of kinematic distributions, which are of central importance for precision
measurements and new physics searches at the LHC and future high-energy colliders.
The present implementation opens up the systematic
study of two-loop EW corrections for a wide range of LHC processes within fully
exclusive Monte Carlo simulations.
 Dedicated studies of remaining uncertainty estimates for important scattering processes,
would be valuable both for precision QCD+EW simulations and for the design of analysis strategies at the
HL-LHC and future high-energy colliders such as the FCC-hh, where EW Sudakov
logarithms are expected to dominate the radiative corrections across large portions
of phase space.

The present implementation is restricted to external massless fermions and transversely
polarised gauge bosons. An important extension is the inclusion of longitudinally
polarised vector bosons, massive fermions, and external Higgs bosons. The inclusion of
longitudinal modes requires a consistent application of the Goldstone-boson
equivalence theorem beyond one loop and a careful treatment of the associated mixing
between longitudinal and Goldstone modes in the broken
theory. These extensions will be presented by us in a future publication.

\section*{Acknowledgments}	
We thank Stefano Pozzorini for invaluable discussions and for comments on the manuscript. 
J.L. is supported by the Science and Technology Research Council (STFC) under the Consolidated Grant ST/X000796/1, and the STFC Ernest Rutherford Fellowship ST/S005048/1. 
The work of L.M. is supported by the Italian Ministry of University and Research (MUR) grant PRIN 2022SNA23K funded by the European Union -- Next Generation EU, Mission 4, Component 2, CUP D53D23002880006.
We acknowledge the use of the DiRAC Cumulus HPC facility under Grant No. PPSP226.

\FloatBarrier

%%%%%%%%%%%%%%%%%%%%%%%%%%%%%%%%%%%%%%%%%%%%%%%
\section{Appendix} \label{appendix}
%%%%%%%%%%%%%%%%%%%%%%%%%%%%%%%%%%%%%%%%%%%%%%%
\appendix
\numberwithin{equation}{section}

%%%%%%%%%%%%%%%%%%%%%%%%%%%%%%%%%%%%%%%%%%%%%%%
\section{List of relevant \OpenLoops parameters for two-loop EW Sudakov logs} \label{app:parameters2loop}
%%%%%%%%%%%%%%%%%%%%%%%%%%%%%%%%%%%%%%%%%%%%%%%

In Tab. \ref{tab:par2} we list input parameters and switches available in \OpenLoops
that are relevant for the selection and evaluation of the two-loop EW corrections in logarithmic approximation. These parameters can be set via the \texttt{set\_parameter} routine in \OpenLoops, see Appendix A.2 in Ref.~\cite{Buccioni:2019sur}. For additional parameters selecting one-loop contributions and settings see Appedix C in \cite{Lindert:2023fcu}.

\begin{table*}[bth!]
\centering
\begin{tabular}{ccl}
\hline
\hline
parameter & type/ default value   &description \\
\hline
\hline
 \multirow{3}{*}{\texttt{nllew}\textunderscore \texttt{on}}  & \multirow{3}{*}{int, default=0} &0: NLO EW \\
 & &1: NLL(') EW \\
 & &2: NLO EW + two-loop  (N)LL EW \\
 & &-2:  two-loop  (N)LL EW \\
\hline
\multirow{4}{*}{\texttt{nllew}\textunderscore \texttt{2l}\textunderscore  \texttt{contrib}}  & \multirow{4}{*}{int, default=0} &  0:  NLL+NLL contribution ($\alpha^2(L^4+L^3)$)\\
 & &1:  LL contribution ($\alpha^2L^4$) \\
 & &2:  NLL a.i. contribution ($\alpha^2L^3$) \\
 & &3:  NLL a.d. contribution ($\alpha^2L^3$) \\
  \hline
 \multirow{2}{*}{\texttt{ew}\textunderscore \texttt{ct}\textunderscore \texttt{b}}  & \multirow{2}{*}{int, default=0} &  0:  $\muren=m_W$  in $\hat\delta^{(2)}_{\mathrm{PR1}}$\\
 & &1: $\muren=\sqrt{s}$  in $\hat\delta^{(2)}_{\mathrm{PR1}}$\\
 \hline
 \multirow{2}{*}{\texttt{ew}\textunderscore \texttt{ct}}  & \multirow{2}{*}{int, default=0} &  0:  $\muren=m_W$  in $\hat\delta^{(2)}_{\mathrm{PR2}}$\\
 & &1: $\muren=\sqrt{s}$  in $\hat\delta^{(2)}_{\mathrm{PR2}}$\\
\hline\hline
\end{tabular}
\caption{Available input parameters and switches that control the contributions of the two-loop EW corrections in logarithmic approximation in \OpenLoops at NLL accuracy. }
\label{tab:par2}
\end{table*}

%%%%%%%%%%%%%%%%%%%%%%%%%%%%%%%%%%%%%%%%%%%%%%%
\section{Explicit factorisable diagrams}\label{app:factorisable_diags}
%%%%%%%%%%%%%%%%%%%%%%%%%%%%%%%%%%%%%%%%%%%%%%%
In the following we report the full list of factorisable diagrams, including their explicit expressions:

\begin{itemize}

\item Two-leg ladder diagrams:
 \begin{align} \label{diag:D1}
&\mathcal{M}_1^{(2), jk} =    \vcenter{\hbox{\begin{tikzpicture}
    \begin{feynman}
      \vertex (a) at ( 1, 0);
      \vertex (c) at ( -1.5, 1.5);
      \vertex (t) at ( 2, 1);
      \vertex (r) at ( 2, -1);
      \vertex[dot] (k) at ( -0.65, 0.65) {\contour{black}{}};
      \vertex[dot] (z) at ( -0.65, -0.65) {\contour{black}{}};
      \vertex[dot] (k1) at ( -1.25, 1.25) {\contour{black}{}};
      \vertex[dot] (z1) at ( -1.25, -1.25) {\contour{black}{}};
      \vertex (v2) at ( -0.95, 0) {\(V_2\)}; 
      \vertex (jname) at ( -1.65, 1.6) {\(j\)}; 
      \vertex (kname) at ( -1.65, -1.6) {\(k\)}; 
      \vertex (d) at ( -1.5, -1.5);
        \vertex[draw,circle,minimum size=0.75cm] (q) at ( 0, 0) {\contour{black}{}};
      \diagram* {
    (q)--[plain] (z),
       (z) -- [plain] (d),
        (q) -- [plain] (k),
       (k) -- [plain]  (c),
       (z) -- [photon]  (k),
       (z1) -- [photon, edge label=\(V_1\)]  (k1)
         };
\end{feynman}
  \end{tikzpicture}}} \\
  &= \mathcal{M}_0 \sum_{V_1, V_2=A,W^\pm,Z} I_j^{\bar V_1} I_j^{\bar V_2} I_k^{V_1} I_k^{V_2} D_1(m_{V_1},m_{V_2}; r_{jk} ) \nonumber
\end{align}

\item Two-leg crossed ladder diagrams:
 \begin{align} \label{diag:D2}
  &\mathcal{M}_2^{(2), jk} =      \vcenter{\hbox{\begin{tikzpicture}
    \begin{feynman}
      \vertex (a) at ( 1, 0);
      \vertex (c) at ( -1.5, 1.5);
      \vertex (t) at ( 2, 1);
      \vertex (r) at ( 2, -1);
      \vertex[dot] (k) at ( -0.75, 0.75) {\contour{black}{}};
      \vertex[dot] (z) at ( -0.75, -0.75) {\contour{black}{}};
      \vertex[dot] (k1) at ( -1.25, 1.25) {\contour{black}{}};
      \vertex[dot] (z1) at ( -1.25, -1.25) {\contour{black}{}};
      \vertex (v1) at ( -1.5, 0.65) {\(V_1\)}; 
     \vertex (v2) at ( -1.5, -0.65) {\(V_2\)}; 
      \vertex (jname) at ( -1.65, 1.6) {\(j\)}; 
      \vertex (kname) at ( -1.65, -1.6) {\(k\)}; 
      \vertex (d) at ( -1.5, -1.5);
        \vertex[draw,circle,minimum size=0.75cm] (q) at ( 0, 0) {\contour{black}{}};
      \diagram* {
    (q)--[plain] (z),
       (z) -- [plain] (d),
        (q) -- [plain] (k),
       (k) -- [plain]  (c),
       (z) -- [photon]  (k1),
       (z1) -- [photon]  (k)
         };
\end{feynman}
  \end{tikzpicture}}}\\
  &= \mathcal{M}_0 \sum_{V_1, V_2=A,W^\pm,Z} I_j^{\bar V_1} I_j^{\bar V_2} I_k^{V_1} I_k^{V_2} D_2(m_{V_1},m_{V_2}; r_{jk} ) \nonumber
\end{align}

\item Two-leg Yang–Mills diagrams: 
 \begin{align} \label{diag:D3}
    &\mathcal{M}_3^{(2), jk} =     \vcenter{\hbox{\begin{tikzpicture}
    \begin{feynman}
      \vertex (a) at ( 1, 0);
      \vertex (c) at ( -1.5, 1.5);
      \vertex (t) at ( 2, 1);
      \vertex (r) at ( 2, -1);
      \vertex[dot] (k) at ( -0.5, 0.5) {\contour{black}{}};
      \vertex[dot] (z) at ( -0.95, -0.95) {\contour{black}{}};
      \vertex[dot] (k1) at ( -1.05, 1.05) {\contour{black}{}};
      \vertex[dot] (z1) at ( -1.5, 0) {\contour{black}{}};
      \vertex (v1) at ( -1.55, 0.8) {\(V_1\)}; 
      \vertex (v2) at ( -0.85, 0) {\(V_2\)}; 
      \vertex (v3) at ( -1.55, -0.8) {\(V_3\)}; 
      \vertex (jname) at ( -1.65, 1.6) {\(j\)}; 
      \vertex (kname) at ( -1.65, -1.6) {\(k\)}; 
      \vertex (d) at ( -1.5, -1.5);
        \vertex[draw,circle,minimum size=0.75cm] (q) at ( 0, 0) {\contour{black}{}};
      \diagram* {
    (q)--[plain] (z),
       (z) -- [plain] (d),
        (q) -- [plain] (k),
       (k) -- [plain]  (c),
       (z) -- [photon]  (z1),
       (k1) -- [photon]  (z1),
      (k) -- [photon]  (z1)
         };
\end{feynman}
  \end{tikzpicture}}}\\
  &= \mathcal{M}_0 \sum_{V_1, V_2, V_3=A,W^\pm,Z} \tilde{\varepsilon}^{V_1 V_2 V_3} I_j^{\bar V_1} I_j^{\bar V_2} I_k^{\bar V_3}  \nonumber \\ &D_3(m_{V_1},m_{V_2},m_{V_3}; r_{jk} ) \nonumber
\end{align}
with 

\begin{equation*}
\tilde{\varepsilon}^{V_1 V_2 V_3} =
\begin{cases}
  (-1)^{p+1} \frac{c_{\text{w}}}{s_{\text{w}}} & \text{if} \quad V_1V_2V_3=\pi(ZW^+W^-), \\[6pt]
 (-1)^p & \text{if} \quad V_1V_2V_3=\pi(AW^+W^-), \\[6pt]
  0 & \text{otherwise}
\end{cases}
\end{equation*}

\item Two-leg inner self energy diagrams :
 \begin{align} \label{diag:D4}
    &\mathcal{M}_4^{(2), jk} =   \vcenter{\hbox{\begin{tikzpicture}
    \begin{feynman}
      \vertex (a) at ( 1, 0);
      \vertex (c) at ( -1.5, 1.5);
      \vertex (t) at ( 2, 1);
      \vertex (r) at ( 2, -1);
      \vertex[dot] (k) at ( -0.45, 0.45) {\contour{black}{}};
      \vertex (z) at ( -0.7, -0.7) {\contour{black}{}};
    \vertex[dot] (z2) at ( -0.9, 0.9) {\contour{black}{}};
      \vertex[dot] (k1) at ( -1.25, 1.25) {\contour{black}{}};
      \vertex[dot] (z1) at ( -1.25, -1.25) {\contour{black}{}};
     \vertex (v2) at ( -0.5, 1.3) {\(V_2\)}; 
      \vertex (d) at ( -1.5, -1.5);
     \vertex (jname) at ( -1.65, 1.6) {\(j\)}; 
      \vertex (kname) at ( -1.65, -1.6) {\(k\)}; 
        \vertex[draw,circle,minimum size=0.75cm] (q) at ( 0, 0) {\contour{black}{}};
      \diagram* {
    (q)--[plain] (z),
       (z) -- [plain] (d),
        (q) -- [plain] (k),
       (k) -- [plain]  (c),
       (z2) -- [photon, half left]  (k),
        (z1) -- [plain]  (q),
       (z1) -- [photon, edge label=\(V_1\)]  (k1)
         };
\end{feynman}
  \end{tikzpicture}}} \\
  &= -\mathcal{M}_0 \sum_{V_1, V_2=A,W^\pm,Z} I_j^{V_1} I_j^{\bar V_2} I_j^{V_2}  I_k^{\bar V_1} D_4(m_{V_1},m_{V_2}; r_{jk} ) \nonumber
\end{align}

\item Two-leg uchiwa diagrams:
 \begin{align} \label{diag:D5}
    &\mathcal{M}_5^{(2), jk} = \vcenter{\hbox{\begin{tikzpicture}
    \begin{feynman}
      \vertex (a) at ( 1, 0);
      \vertex (c) at ( -1.5, 1.5);
      \vertex (t) at ( 2, 1);
      \vertex (r) at ( 2, -1);
      \vertex[dot] (k) at ( -0.45, 0.45) {\contour{black}{}};
      \vertex (z) at ( -0.7, -0.7) {\contour{black}{}};
    \vertex[dot] (z2) at ( -0.9, 0.9) {\contour{black}{}};
      \vertex[dot] (k1) at ( -1.25, 1.25) {\contour{black}{}};
      \vertex[dot] (z1) at ( -0.9, -0.9) {\contour{black}{}};
     \vertex (v2) at ( -0.6, 1.7) {\(V_2\)}; 
      \vertex (d) at ( -1.5, -1.5);
      \vertex (jname) at ( -1.65, 1.6) {\(j\)}; 
      \vertex (kname) at ( -1.65, -1.6) {\(k\)}; 
        \vertex[draw,circle,minimum size=0.75cm] (q) at ( 0, 0) {\contour{black}{}};
      \diagram* {
    (q)--[plain] (z),
       (z) -- [plain] (d),
        (q) -- [plain] (k),
       (k) -- [plain]  (c),
       (k1) -- [photon, half left]  (k),
        (z1) -- [plain]  (q),
       (z1) -- [photon, edge label=\(V_1\)]  (z2)
         };
\end{feynman}
  \end{tikzpicture}}} \\
  &= -\mathcal{M}_0 \sum_{V_1, V_2=A,W^\pm,Z} I_j^{V_1} I_j^{\bar V_2} I_j^{V_2}  I_k^{\bar V_1} D_5(m_{V_1},m_{V_2}; r_{jk} ) \nonumber
\end{align}

\item Two-leg gauge-boson self-energy diagrams:

 \begin{align} \label{diag:D6}
    &\mathcal{M}_{6-11}^{(2), jk} =  \vcenter{\hbox{\begin{tikzpicture}
    \begin{feynman}
      \vertex (a) at ( 1, 0);
      \vertex (c) at ( -1.5, 1.5);
      \vertex (t) at ( 2, 1);
      \vertex (r) at ( 2, -1);
      \vertex[dot] (k) at ( -1, 1) {\contour{black}{}};
      \vertex[dot] (z) at ( -1, -1) {\contour{black}{}};
      \vertex (k1) at ( -1, 0.25) {\contour{black}{}};
      \vertex (z1) at ( -1, -0.35);
      \vertex (d) at ( -1.5, -1.5);
      \vertex (jname) at ( -1.65, 1.6) {\(j\)}; 
      \vertex (kname) at ( -1.65, -1.6) {\(k\)};       
      \vertex[blob,minimum size=0.75cm] (loop) at ( -1,0) {\contour{black}{ }};
        \vertex[draw,circle,minimum size=0.75cm] (q) at ( 0, 0) {\contour{black}{ }};
      \diagram* {
    (q)--[plain] (z),
       (z) -- [plain] (d),
        (q) -- [plain] (k),
       (k) -- [plain]  (c),
       (z) -- [photon, edge label=\(V_2\)]  (z1),
       (k) -- [photon, edge label'=\(V_1\)] (k1)
         };
\end{feynman}
  \end{tikzpicture}}}
\end{align}

where the shaded loop stands for whatever one-loop insertions which is allowed by EW Feynman rules into the propagator of $V_1$. This implies that the shaded blob is not necessarily a one-loop bubble diagram but it can also be a tadpole (e.g. due to 4-point $VVVV$ vertices)

\item Three-leg ladder diagrams:
 \begin{align} \label{diag:D12}
     &\mathcal{M}_{12}^{(2), jkg} =    \vcenter{\hbox{\begin{tikzpicture}
    \begin{feynman}
      \vertex (a) at ( 1, 0);
      \vertex (c) at ( -1.5, 1.5);
      \vertex (t) at ( 2, 1);
      \vertex (r) at ( 2, -1);
      \vertex (a) at ( -1.75, 0);
      \vertex[dot] (k) at ( -1, 1) {\contour{black}{}};
      \vertex[dot] (z) at ( -1, -1) {\contour{black}{}};
      \vertex[dot] (z1) at ( -0.75, 0) {\contour{black}{}};
      \vertex[dot] (z2) at ( -1.5, 0) {\contour{black}{}};
      \vertex (d) at ( -1.5, -1.5);
      \vertex (jname) at ( -1.65, 1.6) {\(k\)}; 
      \vertex (kname) at ( -1.95, 0) {\(j\)};   
      \vertex (kname) at ( -1.65, -1.6) {\(g\)};
      \vertex (v2) at ( -1.2, -0.45) {\(V_2\)};              
        \vertex[draw,circle,minimum size=0.75cm] (q) at ( 0, 0) {\contour{black}{}};
      \diagram* {
    (q)--[plain]  (z),
       (z) -- [plain] (d),
        (q) -- [plain] (k),
       (k) -- [plain]  (c),
       (a) -- [plain]  (q),
      (k) -- [photon, edge label'=\(V_1\)]  (z2),
      (z) -- [photon]  (z1)
         };
\end{feynman}
  \end{tikzpicture}}} \\
  &= \mathcal{M}_0 \sum_{V_1, V_2=A,W^\pm,Z} I_j^{\bar V_1} I_j^{\bar V_2} I_k^{V_2}  I_g^{\bar V_1} D_{12}(m_{V_1},m_{V_2}; r_{jg} ) \nonumber
\end{align}

\item Three-leg Yang–Mills diagrams: 
 \begin{align} \label{diag:D13}
       &\mathcal{M}_{13}^{(2), jkg} =    \vcenter{\hbox{\begin{tikzpicture}
    \begin{feynman}
      \vertex (a) at ( 1, 0);
      \vertex (c) at ( -1.5, 1.5);
      \vertex (t) at ( 2, 1);
      \vertex (r) at ( 2, -1);
      \vertex (a) at ( -1.75, -0.85);
      \vertex[dot] (k) at ( -1, 1) {\contour{black}{}};
      \vertex[dot] (z) at ( -1.25, -1.25) {\contour{black}{}};
      \vertex[dot] (z1) at ( -0.75, -0.38) {\contour{black}{}};
      \vertex[dot] (z2) at ( -1.5, 0) {\contour{black}{}};
      \vertex (jname) at ( -1.65, 1.6) {\(k\)}; 
      \vertex (kname) at ( -2.05, -1) {\(j\)};   
      \vertex (kname) at ( -1.65, -1.6) {\(g\)};            
      \vertex (v2) at ( -1.55, 0.7) {\(V_2\)}; 
      \vertex (v1) at ( -0.95, -0) {\(V_1\)}; 
      \vertex (v3) at ( -1.55, -1.1) {\(V_3\)}; 
      \vertex (d) at ( -1.5, -1.5);
        \vertex[draw,circle,minimum size=0.75cm] (q) at ( 0, 0) {\contour{black}{}};
      \diagram* {
    (q)--[plain]  (z),
       (z) -- [plain]  (d),
        (q) --[plain] (k),
       (k) --[plain]   (c),
       (a) -- [plain]  (q),
      (k) -- [photon]  (z2),
      (z) -- [photon]  (z2),
     (z1) -- [photon]  (z2)
         };
\end{feynman}
  \end{tikzpicture}}}\\
  &= \mathcal{M}_0 \sum_{V_1, V_2=A,W^\pm,Z} \tilde{\varepsilon}^{V_1 V_2 V_3} I_j^{\bar V_1} I_k^{\bar V_2} I_g^{\bar V_3} \nonumber \\ 
  &D_{13}(m_{V_1},m_{V_2},m_{V_3}; r_{jk}, r_{jg}, r_{kg} ) \nonumber
\end{align}

\item Four-leg ladder diagrams: 
 \begin{align} \label{diag:D14}
     &\mathcal{M}_{14}^{(2), jkgh} =       \vcenter{\hbox{\begin{tikzpicture}
    \begin{feynman}
      \vertex (a) at ( 1, 0);
      \vertex (c) at ( -1.5, 1.5);
      \vertex (t) at ( 2, 1);
      \vertex (r) at ( 2, -1);
      \vertex (a) at ( -1.75, 0.65);
      \vertex (b) at ( -1.75, -0.65);
      \vertex[dot] (k) at ( -1, 1) {\contour{black}{}};
      \vertex[dot] (z) at ( -1, -1) {\contour{black}{}};
      \vertex[dot] (z1) at ( -1.2, -0.44) {\contour{black}{}};
      \vertex[dot] (z2) at ( -1.2, 0.44) {\contour{black}{}};
      \vertex (d) at ( -1.5, -1.5);
      \vertex (jname) at ( -1.65, 1.6) {\(j\)}; 
      \vertex (jname) at ( -1.95, 0.75) {\(k\)};  
      \vertex (jname) at ( -1.95, -0.75) {\(g\)};             
      \vertex (kname) at ( -1.65, -1.6) {\(h\)};  
      \vertex (v2) at ( -1.38, -0.9) {\(V_2\)};                     
        \vertex[draw,circle,minimum size=0.75cm] (q) at ( 0, 0) {\contour{black}{}};
      \diagram* {
    (q)--[plain]  (z),
       (z) -- [plain]  (d),
        (q) --[plain]  (k),
       (k) -- [plain]  (c),
       (a) -- [plain]  (q),
       (b) -- [plain]  (q),
       (k) -- [photon, edge label'=\(V_1\)]  (z2),
      (z) -- [photon]  (z1)
         };
\end{feynman}
  \end{tikzpicture}}}\\
  &= \mathcal{M}_0 \sum_{V_1, V_2=A,W^\pm,Z} I_j^{\bar V_1} I_k^{V_1} I_g^{\bar V_2}  I_h^{V_2} D_{14}(m_{V_1},m_{V_2}; r_{jk},r_{gh} ) \nonumber.
\end{align}
\end{itemize}

The explicit expressions at NLL accuracy for the integrals appearing in the diagrams \eqref{diag:D1}--\eqref{diag:D14} are given by 
\begin{align}\label{eq:all_d1_d14_integrals}
 &D_1(m_{V_1},m_{V_2}; r_{jk} ) \overset{\nll}{=}\frac{1}{6} L^4 -\frac{2}{3}\left(2+l_{V_1}-l_{jk}\right)L^3 \nonumber  \\
&D_2 (m_{V_1},m_{V_2}; r_{jk} )  \overset{\nll}{=}\frac{1}{3} L^4 -\frac{4}{3}\left(2+\frac{l_{V_1}+l_{V_2}}{2}-l_{jk}\right)L^3 \nonumber\\ 
&D_3(m_{V_1},m_{V_2},m_{V_3}; r_{jk} )  \overset{\nll}{=}\frac{1}{6} L^4 - \left(3+\frac{l_{V_1}+l_{V_3}}{3}-\frac{2}{3}l_{jk}\right)L^3 \nonumber \\ 
&D_4(m_{V_1},m_{V_2}; r_{jk} )   \overset{\nll}{=}\frac{1}{3} L^3 \nonumber \\ 
&D_5(m_{V_1},m_{V_2}; r_{jk} )   \overset{\nll}{=}-\frac{1}{3} L^3 \nonumber \\ 
&D_{\text{2-loop se}}=\sum_{p=6}^{11} D_{p}(m_{V_1},X_p,m_{V_2}; r_{jk})   \overset{\nll}{\sim} \frac{2}{3} b_{V_1V_2}L^3   \nonumber \\ 
&D_{12}(m_{V_1},m_{V_2}; r_{jg} )   \overset{\nll}{=} \frac{1}{2} L^4 -2\left(2+\frac{2l_{V_1}+l_{V_2}}{3}-l_{jg}\right)L^3 \nonumber \\ 
&D_{13}(m_{V_1},m_{V_2},m_{V_3}; r_{jk}, r_{jg}, r_{kg} ) \overset{\nll}{=} 0 \nonumber \\ 
&D_{14}(m_{V_1},m_{V_2}; r_{jk},r_{gh} ) \overset{\nll}{=} L^4 - 2(2-l_{jk})L^3 - 2(2-l_{gh})L^3
\end{align}
where

\begin{equation}
X_p = \{m_{V_3}\} \qquad \text{or} \qquad X_p = \{m_{V_3}, m_{V_4}\} 
\end{equation}
according to the number of virtual particles in the shaded blob of \eqref{diag:D6}, and $l_V$ is given by
\begin{equation}\label{eq:logMV}
 l_{V_i}:=\log\left(\frac{m_{V_i}}{m_W}\right).
\end{equation}
Note that for $D_{\text{2-loop se}}$ in Eq. \eqref{eq:all_d1_d14_integrals} we used again the symbol "$\sim$", which has the same meaning already explained in the main text in the context of Eq. \eqref{eq:simplified22}. In particular, dropped additional terms exactly cancel against the same opposite sign contribution appearing in the sum of $D_1-D_5$.

Our standard implementation of two-loop soft-collinear EW logarithms in \OpenLoops is based on the combination of these diagrams in terms of the three topologies as given in Eqs. \eqref{eq:simplified22}-\eqref{eq:simplified1111}. However, a flexible separation in terms of the individual diagrams $D_1$--$D_{11}$ is also available.

\bibliographystyle{JHEP}
\bibliography{ewsud2l}

\end{document}